\documentclass[aps,prd,reprint,superscriptaddress,longbibliography,floatfix]{revtex4-2}

\usepackage{amsmath,amssymb}
\usepackage{bm}
\usepackage{graphicx}
\usepackage[T1]{fontenc}
\usepackage{hyperref}
\hypersetup{colorlinks=true,breaklinks=true,linkcolor={[rgb]{0.10,0.25,0.55}},citecolor={[rgb]{0.10,0.25,0.55}},urlcolor={[rgb]{0.10,0.25,0.55}}}

\providecommand{\K}{}\renewcommand{\K}{\mathbf{K}}
\providecommand{\qv}{}\renewcommand{\qv}{\mathbf{q}}
\providecommand{\rvec}{}\renewcommand{\rvec}{\mathbf{r}}
\providecommand{\gv}{}\renewcommand{\gv}{\mathbf{g}}
\providecommand{\Mmat}{}\renewcommand{\Mmat}{\mathbf{M}}
\providecommand{\Pq}{}\renewcommand{\Pq}{\mathbf{P}_{q}}
\providecommand{\khat}{}\renewcommand{\khat}{\hat{\mathbf{k}}}
\providecommand{\abaralpha}{}\renewcommand{\abaralpha}{\bar{\alpha}}

\providecommand{\ULDM}{}\renewcommand{\ULDM}{\mathrm{ULDM}}
\providecommand{\STF}{}\renewcommand{\STF}{\mathrm{STF}}
\providecommand{\qabs}[1]{}\renewcommand{\qabs}[1]{\left|#1\right|}
\providecommand{\qnorm}[1]{}\renewcommand{\qnorm}[1]{\left\|#1\right\|}

\renewcommand{\topfraction}{0.9}
\renewcommand{\bottomfraction}{0.6}
\renewcommand{\textfraction}{0.05}
\renewcommand{\floatpagefraction}{0.85}

\begin{document}
\raggedbottom
\widowpenalty=200 \clubpenalty=200 \displaywidowpenalty=100
\predisplaypenalty=200 \brokenpenalty=100
\renewcommand{\topfraction}{0.9}\renewcommand{\bottomfraction}{0.6}
\renewcommand{\textfraction}{0.05}\renewcommand{\floatpagefraction}{0.85}
\setlength{\parskip}{0pt plus 2pt}
\setlength{\floatsep}{8pt plus 6pt minus 2pt}
\setlength{\textfloatsep}{12pt plus 10pt minus 4pt}
\setlength{\intextsep}{8pt plus 6pt minus 2pt}

\title{Composition-dependent ultralight dark matter detection with a tetrahedral superconducting gravity gradiometer}

\author{John J. Oh}
\email{johnoh@nims.re.kr}
\affiliation{National Institute for Mathematical Sciences, Daejeon 34047, Republic of Korea}
\author{Ho Jung Paik}
\email{hpaik@umd.edu}
\affiliation{Department of Physics, University of Maryland, College Park, MD 20742, USA}

\begin{abstract}
Tetrahedral superconducting gravity gradiometers designed as full-tensor gravitational-wave detectors, such as the ground-based mSOGRO design and the proposed lunar LANGO observatory, reconstruct the acceleration-gradient tensor from four test masses. We show that giving the four test masses two compositions in a $2+2$ arrangement adds a composition-dependent ultralight dark matter (ULDM) channel to that same tensor readout, with no separate dark-matter sensor and no dark-matter-specific readout. A coherently oscillating scalar accelerates each mass in proportion to its dilaton charge, and the tensor reconstruction leaves the charges only in a mean, rejected with the common mode, and a single vector $\qv$ proportional to the charge contrast $\Delta\abaralpha_{AB}$ between the two materials. The response is the rank-one tensor $\K_{\ULDM}(t)=\gv_\phi(t)\otimes\qv$, whose amplitude is set by the halo, independent of $m_\phi$. The axis of $\qv$ is fixed in advance by the material assignment, so the response obeys the null condition $\K_{\ULDM}(\mathbf{I}-\Pq)=0$ at every instant, for any $\gv_\phi$ including a virialized field with a stochastic gradient---an exact tensor discriminator against gravitational-wave signals; for a single composition $\qv=0$. With the gravitational-wave noise model, the coupling sensitivity for copper and lead test masses reaches $d_{\hat m}-d_g\simeq3.4\times10^{-8}$ for the baseline instrument and $4.9\times10^{-10}$ with upgraded accelerometers. The rank-one factorization and the null condition are summarized in a companion Letter; the present paper gives the derivations and the detector-level analysis.
\end{abstract}

\keywords{ultralight dark matter, superconducting gravity gradiometer, equivalence principle, dilaton, gravitational-wave detector}

\maketitle

\section{Introduction}
\label{sec:intro}

If dark matter is a scalar field of low mass, its local realization in the Galactic halo is a classical wave. The occupation number of a mode of mass $m_\phi\lesssim1\,\mathrm{eV}/c^2$ at the solar position is enormous, so the field oscillates coherently at the Compton frequency $f_\phi=m_\phi c^2/h$. The halo has a velocity spread, so modes moving at slightly different speeds oscillate at slightly different frequencies and dephase. The field therefore holds a definite amplitude and phase only for a finite number of cycles, $Q_{\rm coh}=\omega_\phi/\Delta\omega_\phi\sim v_{\rm DM}^{-2}\sim10^{6}$, corresponding to a coherence time $\tau_{\rm coh}\sim Q_{\rm coh}/f_\phi$~\cite{HuBarkana2000,HuiOstriker2017,FreeseLisantiSavage2013}. A field that couples to the Standard Model through a dilaton-like interaction varies the fundamental constants that set atomic and nuclear binding energies, and the variation is composition dependent. Two bodies of different composition acquire accelerations that differ in proportion to their dilaton charges~\cite{DamourDonoghue2010,DamourDonoghueCQG2010,Hees2018}. This composition dependence is the same effect probed statically by tests of the universality of free fall, and static tests already constrain such couplings, while differential accelerometers, torsion balances and dual-species atom interferometers have been proposed as oscillating search platforms~\cite{Touboul2022,Berge2018,Graham2016}.

An oscillating equivalence-principle violation is a natural target for an instrument that already compares the accelerations of separated test masses (TMs). Superconducting gravity gradiometers (SGGs) built from arrays of levitated or suspended TMs were developed for geodesy and, more recently, as full-tensor gravitational-wave detectors~\cite{MoodyChanPaik1986,ChanMoodyPaik1987,Paik2016,Paik2020,Griggs2017}. The tetrahedral form is realized in the ground-based miniaturized SOGRO (mSOGRO) design and proposed for the lunar LANGO observatory~\cite{LANGO2026,Metzler2022}. Four TMs at the vertices reconstruct the full acceleration-gradient tensor, from which a gravitational wave is read out. Such an instrument is, in its usual operation, blind to dark matter. With TMs of a single composition the equivalence-principle-violating acceleration is common to all four masses and is rejected together with every other common-mode disturbance.

This paper shows that the same hardware acquires a channel sensitive to an oscillating scalar field once TMs of more than one composition are introduced. The spatial arrangement of the two compositions then determines the response. Assigning two materials to the four vertices in a $2+2$ pattern
(Fig.~\ref{fig:tetrahedron}) turns $\Delta\abaralpha_{AB}$ into a time-independent vector $\qv$ in the reconstructed tensor. The dark-matter response becomes the rank-one tensor
$\K_{\ULDM}(t)=\gv_\phi(t)\otimes\qv$, with $\gv_\phi$ the gradient of the scalar field. The proposed design of the TMs with two different compositions is given in Appendix~\ref{app:tmdesign}. The time dependence comes entirely from that gradient, and the axis of $\qv$ is set by the material assignment and the geometry. No dark-matter sensor and no dark-matter-specific readout are added; the signal appears in the channels that already carry gravitational waves, distinguished from them by a null condition that follows from the factorization. The vector $\qv$ is real, and the axis it defines is fixed once the materials are assigned. The reconstructed tensor is therefore confined at every instant to the matrices with every row a multiple of $\qv^{\rm T}$, so that
\begin{equation}
\K_{\ULDM}(\mathbf{I}-\Pq)=0,
\qquad
\Pq=\frac{\qv\,\qv^{\rm T}}{\qabs{\qv}^{2}} .
\label{eq:null-intro}
\end{equation}
The condition constrains $\qv$ alone and therefore holds for any $\gv_\phi$. In a virialized halo, the gradient does not oscillate along a fixed direction but traces an ellipse, and Eq.~(\ref{eq:null-intro}) holds for each instantaneous value. This condition separates the dark-matter response from a gravitational or a rotational background taken alone, and Sec.~\ref{sec:null} treats the correlated combinations.

The rank-one factorization and the null condition it implies are summarized in a companion Letter~\cite{companionLetter}. The present paper gives the derivations and the detector-level analysis. The remainder of the paper develops this picture and its consequences. Section~\ref{sec:signal} sets out the classical ULDM field and its composition-dependent acceleration, including the stochastic form of the gradient. Section~\ref{sec:reconstruction} reviews the tetrahedral reconstruction and the inertial corrections specific to a rotating, suspended platform. Section~\ref{sec:response} derives the rank-one response and the null condition~(\ref{eq:null-intro}), and identifies what survives when the field is stochastic. Section~\ref{sec:assignment} treats the composition assignment and shows that the six $2+2$ splits are equivalent and fixes the composition axis. Section~\ref{sec:sensitivity} discusses the coupling sensitivity of the configuration and the gain from operating two tetrahedra on a common frame. Section~\ref{sec:discussion} summarizes the detection principle and its relation to existing composition-dependent tests.

\section{The ULDM signal}\label{sec:signal}

\subsection{Coherent field and its gradient}

A real scalar field $\phi$ of mass $m_\phi$ minimally coupled to gravity obeys the Klein--Gordon equation, whose non-relativistic solution at the solar position is a traveling wave
\begin{equation}
\phi(t,\rvec)=\phi_0\cos(\omega_\phi t-\bm k_\phi\!\cdot\!\rvec+\theta),
\label{eq:planewave}
\end{equation}
with dispersion $\omega_\phi^2=m_\phi^2+\lvert\bm k_\phi\rvert^2$, where $\bm k_\phi$ is the wavevector and $\theta$ a constant phase.
For a virial velocity $v_{\rm DM}\simeq10^{-3}c$ the field is strongly non-relativistic, $\lvert\bm k_\phi\rvert\simeq m_\phi v_{\rm DM}$ and $\omega_\phi\simeq m_\phi c^2/\hbar$, so it oscillates at
\begin{equation}
f_\phi=\frac{m_\phi c^2}{h}
\simeq 2.4\times10^{2}\,\mathrm{Hz}\left(\frac{m_\phi c^2}{10^{-12}\,\mathrm{eV}}\right),
\label{eq:fphi}
\end{equation}
placing the mass window $m_\phi c^2=10^{-18}$--$10^{-15}\,\mathrm{eV}$ of interest here at $f_\phi\simeq0.24\,\mathrm{mHz}$--$0.24\,\mathrm{Hz}$, within the band of a terrestrial gradiometer. The de Broglie wavelength $\lambda_{\rm dB}=2\pi\hbar/(m_\phi v_{\rm DM})$ runs from $10^{12}$ to $10^{15}\,$m across the mass window, more than eleven orders of magnitude above any laboratory baseline. For $L=0.4\,$m the phase difference across the array is $\lvert\bm k_\phi\rvert L\lesssim2\times10^{-12}$, so all four TMs see one and the same field gradient. A differential signal therefore cannot arise from a phase difference between the masses. It can arise only from a difference in composition between the masses. The occupation of a de Broglie volume, $\mathcal N\sim(\rho_{\rm DM}/m_\phi c^2)\lambda_{\rm dB}^3\sim10^{66}$ or larger, allows the field to be treated as a classical wave rather than as a quantum operator~\cite{HuBarkana2000,HuiOstriker2017}.

The amplitude follows from the local density. Averaging the field energy density over an oscillation gives $\rho_\phi\simeq\tfrac12 m_\phi^2\phi_0^2$, so that with $\rho_\phi=\rho_{\rm DM}$
\begin{equation}
\phi_0=\frac{\sqrt{2\rho_{\rm DM}}}{m_\phi},~~~
\lvert\nabla\phi\rvert\simeq v_{\rm DM}\sqrt{2\rho_{\rm DM}} ,
\label{eq:phi0}
\end{equation}
the gradient amplitude following from $\lvert\nabla\phi\rvert=\lvert\bm k_\phi\rvert\phi_0$.
Throughout, $\rho_{\rm DM}=0.4\,\mathrm{GeV\,cm^{-3}}$~\cite{Read2014} is used in its mass-density form $\rho_{\rm DM}/c^2=7.1\times10^{-22}\,\mathrm{kg\,m^{-3}}$ wherever it enters an SI expression. The fractional frequency width $\Delta\omega_\phi/\omega_\phi\sim v_{\rm DM}^2\sim10^{-6}$ fixes a coherence time $\tau_{\rm coh}\sim Q_{\rm coh}/f_\phi$ with $Q_{\rm coh}\sim10^6$; the signal is therefore a narrow line, resolvable in frequency throughout the band. The coherence time exceeds a sidereal day for $f_\phi\lesssim12\,\mathrm{Hz}$ and a year-long observation for $f_\phi\lesssim30\,\mathrm{mHz}$, so over most of the mass window such an observation is a single coherence segment.

The single mode of Eq.~(\ref{eq:planewave}) is an idealization. A virialized halo is a superposition of modes with wavevectors in different directions, so within one coherence interval the physical field gradient is
\begin{equation}
\gv_\phi(t)=\bm u\cos\omega_\phi t+\bm v\sin\omega_\phi t ,
\label{eq:stochgrad}
\end{equation}
with $\bm u$ and $\bm v$ two real vectors that are in general not parallel. The tip of $\gv_\phi(t)$ therefore traces an ellipse in the plane spanned by $\bm u$ and $\bm v$. The ellipse degenerates to a line when $\bm u$ and $\bm v$ are parallel, and $\gv_\phi$ then reduces to $-A_0\khat\cos(\omega_\phi t+\theta)$, which is the single-mode case. Each mode contributes along its own wavevector with its own phase, which is why the two vectors are not parallel in general. Section~\ref{sec:response} carries this character through to the response.

\subsection{Dilatonic coupling and the composition-dependent acceleration}

A dilaton-like scalar couples to the electromagnetic, gluonic and fermion-mass sectors of the Standard Model through~\cite{DamourDonoghue2010,DamourDonoghueCQG2010}
\begin{equation}
\begin{aligned}
\mathcal L_{\rm int}=\varphi\bigg[&\frac{d_e}{4e^2}F_{\mu\nu}F^{\mu\nu}
-\frac{d_g\beta_3}{2g_3}G^A_{\mu\nu}G_A^{\mu\nu}\\
&-\sum_i\!\left(d_{m_i}+\gamma_{m_i}d_g\right)m_i\bar\psi_i\psi_i\bigg],
\end{aligned}
\label{eq:Lint}
\end{equation}
where $F_{\mu\nu}$ and $G^A_{\mu\nu}$ are the electromagnetic and gluon field-strength tensors, $\psi_i$ and $m_i$ the fermion fields and their masses, $e$ and $g_3$ the gauge couplings, $\beta_3$ the QCD beta function and $\gamma_{m_i}$ the mass anomalous dimensions. The five independent couplings are
\begin{equation}
c\in\{e,\;g,\;\hat m,\;\delta m,\;m_e\},
\label{eq:couplings}
\end{equation}
the electromagnetic and gluon field strengths, the isospin-symmetric light-quark mass, the light-quark mass difference and the electron mass, and the field is normalized as $\varphi=\kappa\phi$ with $\kappa=\sqrt{4\pi G}=M_*^{-1}$, $M_*=(4\pi G)^{-1/2}=\sqrt2\,\bar M_{\rm Pl}$.

The instrument responds to the force on a macroscopic TM, which enters through the field dependence of its rest mass. We define the scalar charge as the logarithmic derivative
$\alpha_A\equiv\partial\ln m_A(\varphi)/\partial\varphi$, so that
$m_A(\varphi)=m_A^{(0)}[1+\alpha_A\varphi+\mathcal O(\varphi^2)]$. Transferring the field
dependence onto the fine-structure constant, the QCD scale and the fermion masses, Damour and
Donoghue obtain~\cite{DamourDonoghue2010,Hees2018}
\begin{equation}
\begin{aligned}
\alpha_A=d_g&+(d_{\hat m}-d_g)Q_{\hat m,A}+(d_{\delta m}-d_g)Q_{\delta m,A}\\
&+(d_{m_e}-d_g)Q_{m_e,A}+d_e Q_{e,A},
\end{aligned}
\label{eq:alphaA}
\end{equation}
where the $Q_{c,A}$ are the material-dependent dilaton charges of body $A$, collected with their numerical values in Appendix~\ref{app:charges}. For the metals used here the combination is dominated by the $(d_{\hat m}-d_g)$ and $d_e$ terms. The field-dependent rest mass acts as a potential energy for the body, and its gradient gives the acceleration
\begin{equation}
\bm a_A^\varphi(t)=-\kappa\,\alpha_A\,\nabla\phi(t)=\alpha_A\,\gv_\phi(t),
\label{eq:aA}
\end{equation}
where the characteristic amplitude is
\begin{equation}
\begin{aligned}
A_0&\equiv\kappa\lvert\nabla\phi\rvert
=\kappa\lvert\bm k_\phi\rvert\phi_0
=\sqrt{4\pi G}\,(m_\phi v_{\rm DM})\frac{\sqrt{2\rho_{\rm DM}}}{m_\phi}\\
&=\sqrt{8\pi G\rho_{\rm DM}}\;v_{\rm DM}
=2.95\times10^{-10}\,\mathrm{m\,s^{-2}} ,
\end{aligned}
\label{eq:A0}
\end{equation}
in units with $\hbar=c=1$. The values are $\rho_{\rm DM}=0.4\,\mathrm{GeV}/c^2\,\mathrm{cm^{-3}}$ and $v_{\rm DM}=\sqrt{3/2}\,v_0=269\,\mathrm{km\,s^{-1}}$, where $v_0=220\,\mathrm{km\,s^{-1}}$ is the local circular speed, that of a circular orbit at the solar radius~\cite{FreeseLisantiSavage2013}. The order-of-magnitude value $v_{\rm DM}\simeq10^{-3}c$ used above for the kinematics is not accurate enough here, because $A_0$ is linear in $v_{\rm DM}$. Each mass is then pushed along the common field gradient $\gv_\phi$ with a strength set by its own charge. The scalar mass cancels, so $A_0$ is set by the halo alone, not by $m_\phi$, the TM or the baseline. The composition enters through $\alpha_A$ instead.

The charge is defined for a single nuclide, whereas a TM may be an alloy or a combination of multiple elements. So a rule is needed for combining constituents. Newton's second law gives $M\bm a_{\rm cm}=\sum_i m_i\bm a_i$, so each is weighted by its own mass and the relevant quantity is the mass-weighted mean charge,
\begin{equation}
\abaralpha_A=\frac{\sum_i m_i\alpha_i}{\sum_i m_i},
\label{eq:abar}
\end{equation}
and the differential acceleration between two materials is
\begin{equation}
\Delta\bm a_{AB}^\varphi(t)=\Delta\abaralpha_{AB}\,\gv_\phi(t),
~~~
\Delta\abaralpha_{AB}\equiv\abaralpha_A-\abaralpha_B .
\label{eq:diffacc}
\end{equation}
The common part $\abaralpha_0$ pushes all four masses alike and is rejected with the translational common mode, so only $\Delta\abaralpha_{AB}$ reaches the tensor channels. The next section shows how the tetrahedral geometry converts it into a fixed axis in the reconstructed tensor.

\section{Tetrahedral reconstruction}\label{sec:reconstruction}

\subsection{Geometry and the reconstruction formula}

The four TMs are placed at the vertices of a regular tetrahedron, at positions
\begin{equation}
\begin{aligned}
\rvec_1&=\tfrac{L}{2\sqrt2}(1,1,1)^{\rm T}, & \rvec_2&=\tfrac{L}{2\sqrt2}(-1,1,-1)^{\rm T},\\ \rvec_3&=\tfrac{L}{2\sqrt2}(1,-1,-1)^{\rm T}, & \rvec_4&=\tfrac{L}{2\sqrt2}(-1,-1,1)^{\rm T} .
\end{aligned}
\label{eq:vertices}
\end{equation}
The vertices are centered, $\sum_{i=1}^4\rvec_i=0$. Their second-moment matrix is
\begin{equation}
\Mmat\equiv\sum_{i=1}^4\rvec_i\rvec_i^{\rm T} =\frac{L^2}{2}\,\mathbf I .
\label{eq:Mmatrix}
\end{equation}
The matrix is isotropic: the four vertices occupy alternate corners of a cube of side $L/\sqrt2$, all six vertex pairs are separated by $L$, and no direction is preferred.

\begin{figure*}[t]
\centering
\includegraphics[width=\textwidth]{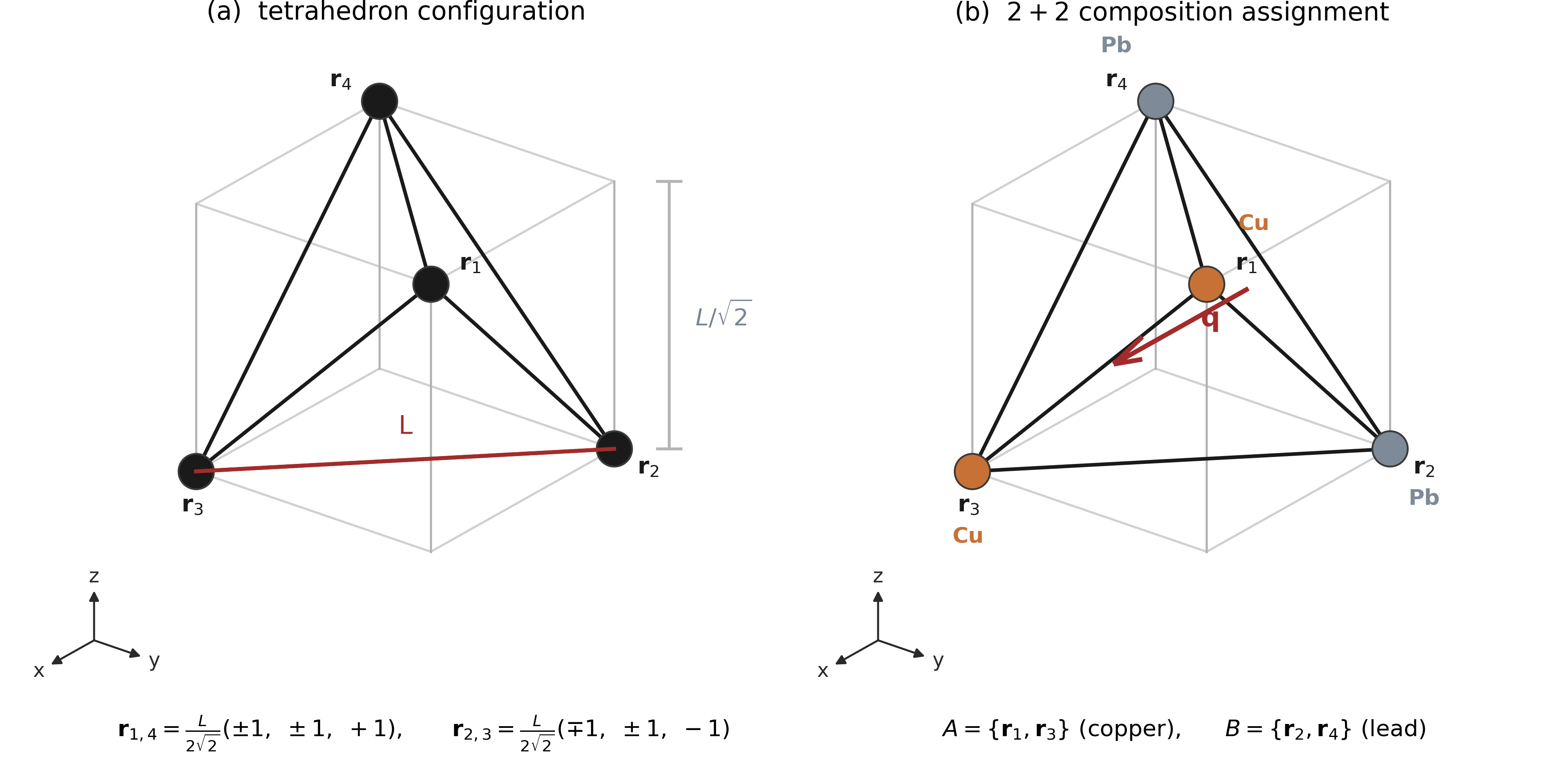}
\caption{\label{fig:tetrahedron}(a) The regular tetrahedron of Eq.~(\ref{eq:vertices}), on alternate corners of a cube of side $L/\sqrt2$, all six vertex pairs separated by $L$. The vertices are centered, $\sum_i\rvec_i=0$, and the second-moment matrix is isotropic, $\Mmat=\tfrac12L^2\mathbf I$. (b) A $2+2$ composition assignment: copper on $\{\rvec_1,\rvec_3\}$ and lead on $\{\rvec_2,\rvec_4\}$, which puts $\qv$ along $\hat{\bm x}$. The other two $2+2$ splits put $\qv$ along $\hat{\bm y}$ and $\hat{\bm z}$ and are equivalent to this one, since $\Mmat$ is isotropic.}
\end{figure*}

The array is small compared with the scale on which the acceleration varies, so at each mass the acceleration is a constant plus a term linear in position,
\begin{equation}
\bm a_i(t)=\bm A(t)+\K(t)\rvec_i+\mathcal O(R^2\nabla^2\bm a),
\label{eq:affine}
\end{equation}
with $\bm A$ the common translational acceleration and $\K$ the acceleration-gradient tensor. Averaging over the four masses and using $\sum_i\rvec_i=0$ isolates the common mode,
\begin{equation}
\bm A=\frac14\sum_{i=1}^4\bm a_i,
\label{eq:Avec}
\end{equation}
and subtracting it leaves $\bm a_i-\bm A=\K\rvec_i$. Multiplying on the right by $\rvec_i^{\rm T}$, summing and multiplying by $\Mmat^{-1}$ gives the reconstruction. The common mode drops out of the sum, since the vertices are centered and $\sum_i\bm A\rvec_i^{\rm T}=0$, leaving
\begin{equation}
\K=\left[\sum_{i=1}^4\bm a_i\rvec_i^{\rm T}\right]\Mmat^{-1},
\label{eq:Kreco}
\end{equation}
with
\begin{equation}
\Mmat^{-1}=\frac{2}{L^2}\,\mathbf I .
\label{eq:Minv}
\end{equation}
The three translational and nine tensor components together make twelve channels.

\subsection{Irreducible decomposition}

The nine components of $\K$ split under rotations into a trace, a symmetric-trace-free (STF) part and an antisymmetric part,
\begin{equation}
\begin{aligned}
\K=\tfrac13(\operatorname{Tr}\K)\mathbf I
&+\left[\tfrac12(\K+\K^{\rm T})-\tfrac13(\operatorname{Tr}\K)\mathbf I\right]\\
&+\tfrac12(\K-\K^{\rm T}),
\end{aligned}
\label{eq:Kdecomp}
\end{equation}
that is,
\begin{equation}
\begin{aligned}
\K_{\rm tr}&=\tfrac13(\operatorname{Tr}\K)\mathbf I,\\
\K_{\rm STF}&=\tfrac12(\K+\K^{\rm T})-\tfrac13(\operatorname{Tr}\K)\mathbf I,\\
\K_{\rm A}&=\tfrac12(\K-\K^{\rm T}),
\end{aligned}
\label{eq:Kparts}
\end{equation}
with dimensions $1+5+3=9$. This count assumes that the acceleration of each mass is measured in all
three directions. Adding the three translations to the nine tensor components, the channel budget of that ideal readout is
\begin{equation}
12=3_{\rm trans}+3_{\rm rot}+5_{\rm STF}+1_{\rm trace}.
\label{eq:budget}
\end{equation}

\subsection{Gravity and platform rotation}

An ordinary Newtonian field $\bm g=-\nabla\Phi$ enters through the gravity-gradient tensor
\begin{equation}
\Gamma_{ij}=\partial_j g_i=-\partial_i\partial_j\Phi,
\label{eq:Gamma}
\end{equation}
which is symmetric, $\Gamma_{ij}=\Gamma_{ji}$, and traceless in vacuum, $\operatorname{Tr}\bm\Gamma=-\nabla^2\Phi=0$. The geodesic-deviation term of a gravitational wave, $\tfrac12\ddot{\bm h}$, is likewise symmetric and traceless. Both therefore reside entirely in the STF sector; neither contributes a trace or an antisymmetric part.

The antisymmetric sector is generated instead by rotation of the platform. In a frame rotating with angular velocity $\bm\Omega$ the reconstructed tensor acquires the Euler and centrifugal terms
\begin{equation}
\K_{\rm Euler}=\begin{pmatrix}0&\dot\Omega_z&-\dot\Omega_y\\-\dot\Omega_z&0&\dot\Omega_x\\\dot\Omega_y&-\dot\Omega_x&0\end{pmatrix},
\label{eq:Euler}
\end{equation}
\begin{equation}
\K_{\rm cent}=\begin{pmatrix}
\Omega_y^2+\Omega_z^2&-\Omega_x\Omega_y&-\Omega_x\Omega_z\\
-\Omega_x\Omega_y&\Omega_x^2+\Omega_z^2&-\Omega_y\Omega_z\\
-\Omega_x\Omega_z&-\Omega_y\Omega_z&\Omega_x^2+\Omega_y^2
\end{pmatrix}.
\label{eq:cent}
\end{equation}
The Euler term is purely antisymmetric, and its entries are the components of $\dot{\bm\Omega}$, so it measures the angular \emph{acceleration}: $\dot\Omega_z$ is the yaw about the vertical, $\dot\Omega_x$ and $\dot\Omega_y$ the tilts. The centrifugal term is symmetric and does contaminate the gradient.

Only the Euler term has to be kept. Its yaw component $\dot\Omega_z$ lies in the $xy$ block. The two tilt components are the ones the single-cable suspension suppresses.

The centrifugal and Coriolis terms are excluded in a different way. The centrifugal term built from the Earth rate alone is a constant matrix at zero frequency, outside the measurement band, and the cross term with the platform motion has rank two or three. The Coriolis operator $-2[\bm\Omega]_\times$ is antisymmetric. The tensor it leaves in the reconstruction is not. That tensor is $\K_{\rm Cor}=-2[\bm\Omega]_\times V$, with $V$ the velocity gradient of the masses across the array. For a rigid platform $V=[\bm\omega_p]_\times$ and $\K_{\rm Cor}=2(\bm\Omega\cdot\bm\omega_p)\,\mathbf I-2\,\bm\omega_p\bm\Omega^{\rm T}$. The rank-one part has right factor $\hat{\bm\Omega}$, fixed by the sky. The response has right factor $\qv$, fixed by the assignment. The projector, to be discussed in Sec.~\ref{sec:null}, separates the two unless $\hat\qv\parallel\hat{\bm\Omega}$, an alignment the assignment avoids. If the masses move nonrigidly on their suspensions the rank argument does not apply, but the residual Coriolis terms are then suppressed by $\Omega_E$ relative to the band frequencies. Section~\ref{sec:null} uses these structures.

\subsection{Sensing all three degrees of freedom}\label{sec:pendulum}

The twelve channels assume that the acceleration of each mass is measured in
all three directions with equal sensitivity. A mass hung on a wire is soft in
its two horizontal directions but stiff in the vertical, and that stiffness
has to be removed. First, the wire is replaced by a soft spring. We then apply a superconducting negative spring to bring down the vertical resonance frequency to near zero.

The superconducting negative-spring technique is well established. Paik's
group demonstrated it~\cite{Parke1984,MoodyChanPaik1986} and applied an
improved version to their Model-3 SGG~\cite{Paik1996}. 
The mSOGRO design already uses superconducting negative springs to reduce the
horizontal resonance frequencies. Here we apply one along the vertical axis as
well. See Appendix~\ref{app:tmdesign} for the design of the tensor ULDM detector: two TMs of copper and two of lead, with superconducting negative springs applied in both the horizontal and vertical directions.

With the vertical mode softened, the motion of each TM is read in all three directions and the reconstruction of Eq.~(\ref{eq:Kreco})
returns all nine components of $\K$. Gradient and rotation then separate exactly,
\begin{equation}
\begin{aligned}
\Gamma_{xz}&=\tfrac12(K_{xz}+K_{zx}), &\dot\Omega_y&=\tfrac12(K_{zx}-K_{xz}),\\
\Gamma_{yz}&=\tfrac12(K_{yz}+K_{zy}), &\dot\Omega_x&=\tfrac12(K_{yz}-K_{zy}),
\end{aligned}
\label{eq:tiltsolved}
\end{equation}
so the tilts are measured rather than assumed small, and the trace
\begin{equation}
\operatorname{Tr}\K=K_{xx}+K_{yy}+K_{zz}
\label{eq:tracefull}
\end{equation}
is available. The channel budget is that of Eq.~(\ref{eq:budget}).

\section{Composition response and the null condition}\label{sec:response}

\subsection{The rank-one response}

Equation~(\ref{eq:Kreco}) is linear, so the accelerations add and the tensor splits into
\begin{equation}
\K=\bm\Gamma+\K_{\rm Euler}+\K_{\rm cent}+\K_{\ULDM}+\mathbf n .
\label{eq:Ksplit}
\end{equation}
with $\mathbf n$ the instrument noise. The first three terms were treated in Sec.~\ref{sec:reconstruction}. This section covers the fourth term, $\K_{\ULDM}$, obtained by inserting the dark-matter accelerations alone into Eq.~(\ref{eq:Kreco}).

Each TM is pushed along the common field gradient with a strength set by its own charge, so by Eq.~(\ref{eq:aA})
\begin{equation}
\bm a_i^{\ULDM}(t)=\abaralpha_i\,\gv_\phi(t),
\label{eq:ai}
\end{equation}
where $\abaralpha_i$ is the mass-weighted charge of Eq.~(\ref{eq:abar}) for the material at vertex $i$. Inserting this into the reconstruction (\ref{eq:Kreco}) and using the linearity of the sum,
\begin{equation}
\begin{aligned}
\K_{\ULDM}(t)&=\left[\sum_{i=1}^4\abaralpha_i\,\gv_\phi(t)\,\rvec_i^{\rm T}\right]\Mmat^{-1}\\
&=\gv_\phi(t)\left[\Mmat^{-1}\sum_{i=1}^4\abaralpha_i\rvec_i\right]^{\rm T},
\end{aligned}
\label{eq:Kfactor}
\end{equation}
since $\gv_\phi$ does not depend on $i$ and $\Mmat^{-1}$ is symmetric. Defining the \emph{composition vector}
\begin{equation}
\qv\equiv\Mmat^{-1}\sum_{i=1}^{4}\abaralpha_i\rvec_i ,
\label{eq:qdef}
\end{equation}
the response is the outer product
\begin{equation}
\K_{\ULDM}(t)=\gv_\phi(t)\otimes\qv ,
\label{eq:Kuldm}
\end{equation}
a tensor of rank one at every instant, with components $K_{ij}=g_iq_j$. We call $\gv_\phi$ the field gradient vector and $\qv$ the composition vector. The factorization is exact in this common-gradient limit, with finite-wavelength corrections entering at $O(\lvert\bm k_\phi\rvert L)$. The field gradient vector carries the entire time dependence and is common to all four masses. The composition vector carries the entire composition dependence. Its axis is fixed by the material assignment and the geometry, while its magnitude carries the unknown $\Delta\abaralpha_{AB}$. Labeling the two materials so that $\Delta\abaralpha_{AB}>0$ fixes the sense of $\qv$ by convention, and the figures are drawn that way. Swapping the two materials between the groups sends $\qv\to-\qv$ and leaves every composition-independent background as it is. The sense itself is degenerate with the unknown phase of the oscillation and is not observable; the relative sense between two arrays is observable, and Sec.~\ref{sec:null} uses it. For a single plane-wave mode $\gv_\phi(t)=-A_0\khat\cos(\omega_\phi t+\theta)$, Eq.~(\ref{eq:Kuldm}) reduces to
\begin{equation}
\K_{\ULDM}(t)=-A_0\cos(\omega_\phi t+\theta)\,\khat\otimes\qv .
\label{eq:Kuldm-single}
\end{equation}

The reconstructed $\K$ is not the gradient of a single acceleration field but an effective tensor built from four masses responding with different strengths to one common field, so tracelessness does not constrain it.

Under Eq.~(\ref{eq:Kparts}), the outer product fills all three sectors of the tensor, with squared magnitudes
\begin{equation}
\begin{aligned}
\lVert\K_{\rm tr}\rVert^2&=\tfrac13(\gv_\phi\!\cdot\!\qv)^2,\\
\lVert\K_{\rm A}\rVert^2&=\tfrac12\lvert\gv_\phi\!\times\!\qv\rvert^2,\\
\lVert\K_{\STF}\rVert^2&=\tfrac12\lvert\gv_\phi\rvert^2\lvert\qv\rvert^2+\tfrac16(\gv_\phi\!\cdot\!\qv)^2,
\end{aligned}
\label{eq:norms}
\end{equation}
which sum to $\lvert\gv_\phi\rvert^2\lvert\qv\rvert^2$. These are algebraic identities, true at every instant for any $\gv_\phi$, so the three are locked to one another and none is a free parameter.

\subsection{The mean charge and the deviations}

We split the charges into a mean and a deviation, $\abaralpha_i=\abaralpha_0+\delta\abaralpha_i$ with $\abaralpha_0=\tfrac14\sum_i\abaralpha_i$. The deviation is linear across the array, $\delta\abaralpha_i=\qv\cdot\rvec_i$, and this affine structure holds exactly for four masses. The mean part
contributes
\begin{equation}
\qv\big|_{\rm mean}=\abaralpha_0\,\Mmat^{-1}\sum_{i=1}^4\rvec_i=0
\label{eq:qmean}
\end{equation}
because the vertices are centered. The mean charge appears instead in the translational common mode,
\begin{equation}
\bm A_{\ULDM}(t)=\abaralpha_0\,\gv_\phi(t),
\label{eq:Auldm}
\end{equation}
which any gradiometer rejects along with platform acceleration. Only the deviations from the mean survive:
\begin{equation}
\qv=\Mmat^{-1}\sum_{i=1}^{4}\delta\abaralpha_i\,\rvec_i .
\label{eq:qdev}
\end{equation}
In particular, for TMs of a single composition, every $\delta\abaralpha_i$ vanishes and
\begin{equation}
\qv=0
\label{eq:qzero}
\end{equation}
identically. With a single composition, the response is absent from the tensor channels altogether whatever the material and however strong the coupling; with two it is present with a strength proportional to $\Delta\abaralpha_{AB}$.

\subsection{The null condition}\label{sec:null}

The three sectors are not independent. They descend from a single outer product. Since every row of $\gv_\phi\otimes\qv$ is a multiple of $\qv^{\rm T}$, the tensor annihilates every direction orthogonal to $\qv$,
\begin{equation}
\K_{\ULDM}\left(\mathbf I-\Pq\right)=0 ,
\qquad
\Pq=\frac{\qv\,\qv^{\rm T}}{\qabs{\qv}^{2}},
\label{eq:null}
\end{equation}
where $\Pq$ projects onto $\qv$. This imposes six independent conditions on the nine components. The magnitude $\qabs{\qv}\propto\Delta\abaralpha_{AB}$ carries the unknown couplings, but the projector $\Pq$ depends only on the direction $\hat{\qv}$, which Eq.~(\ref{eq:qdef}) fixes from the material assignment and the geometry.

The null condition says that one vector reproduces the whole measured tensor, $\K=(\K\hat{\qv})\otimes\hat{\qv}$. If the measured tensor is $\K_{\ULDM}$, that vector is
\begin{equation}
\K_{\ULDM}(t)\,\hat{\qv}=\qabs{\qv}\,\gv_\phi(t) ,
\label{eq:grecover}
\end{equation}
so the field gradient is recovered up to the factor $\qabs{\qv}$. No nonzero gravity gradient in vacuum can satisfy Eq.~(\ref{eq:null}). If $\K=\bm u\otimes\qv$ is symmetric then $u_iq_j=u_jq_i$, so $\bm u\parallel\qv$ and $\K=\lambda\,\qv\otimes\qv$; tracelessness then forces $\lambda=0$. The same holds for the tidal tensor of a gravitational wave. A platform rotation cannot satisfy it either, since a real antisymmetric $3\times3$ matrix has rank zero or two, never one, and the centrifugal cross term has rank two or three. Any field common to the array that couples through a composition-dependent charge shares the signal structure and passes by construction.

One combination is left. For $\bm u\perp\qv$ the symmetric and antisymmetric parts of $\bm u\otimes\qv$ are an admissible gravity gradient and an admissible Euler term, so a gradient and an angular acceleration locked to the same outer product would pass, and a single array cannot separate them from a transverse signal. Three things break the coincidence. The combination is traceless, whereas the response has $\operatorname{Tr}\K_{\ULDM}=\gv_\phi\cdot\qv$. It would have to persist against the spectral, sidereal and material-exchange tests below. And the difference of the two arrays, Eq.~(\ref{eq:refmode}), cancels the common gradient and rotation to linear order in the gradient; it is the primary statistic of the search. The condition is therefore applied not to the raw record $\K_{\rm meas}=\Gamma+\K_{\rm rot}+\gv_\phi\qv^{\rm T}+\bm n$ but to the candidate component estimated with $\Gamma$ and $\K_{\rm rot}$ as nuisance processes. A candidate that fails under every admissible background is excluded as dark matter; passing does not establish it.

\subsection{The signal model for a virialized halo}

Equation~(\ref{eq:null}) constrains $\qv$ alone and leaves $\gv_\phi$ free, so a virialized halo changes the signal model without touching the condition. Within one coherence interval the gradient traces an ellipse, Eq.~(\ref{eq:stochgrad}), and different rows of $\K$ carry different combinations of $\bm u$ and $\bm v$. One amplitude, one phase and one direction are therefore not enough. A coherence segment needs the six components of $\bm u$ and $\bm v$, and over many segments these are treated as a complex Gaussian vector with a covariance fixed by the halo rather than as parameters to be fitted~\cite{LisantiMoschellaTerrano2021,Centers2021}.

The composition vector is fixed in the laboratory, while $\bm u$ and $\bm v$ are fixed in the Galactic frame over a coherence time, which exceeds a sidereal day over the entire measurement band. The line amplitude in each channel therefore varies over a sidereal day. The variation is a consistency requirement rather than a filter: one pair $(\bm u,\bm v)$ must reproduce the amplitude and phase of every channel across the whole day, and a year of data gives at most eight coherence segments at the top of the band and one below $30\,$mHz. 

\section{The composition assignment}\label{sec:assignment}

\subsection{The $2+2$ assignment}

There are six $2+2$ assignments of the two materials among the four TMs, falling into three pairs related by material exchange. Since
$\Mmat$ is isotropic no direction is suppressed, and each pair defines one of
the three cube axes,
\begin{equation}
\qv=\pm\frac{\Delta\abaralpha_{AB}}{L_{\rm eff}}\,\hat{\bm n},
\qquad \hat{\bm n}\in\{\hat{\bm x},\hat{\bm y},\hat{\bm z}\},
\qquad L_{\rm eff}\equiv\frac{L}{\sqrt2},
\label{eq:qsplits}
\end{equation}
so the three choices are equivalent up to a rotation of the array. The
composition axis is free, and the sign within a pair can change by the material exchange.

\begin{figure}[t]
\centering
\includegraphics[width=\columnwidth]{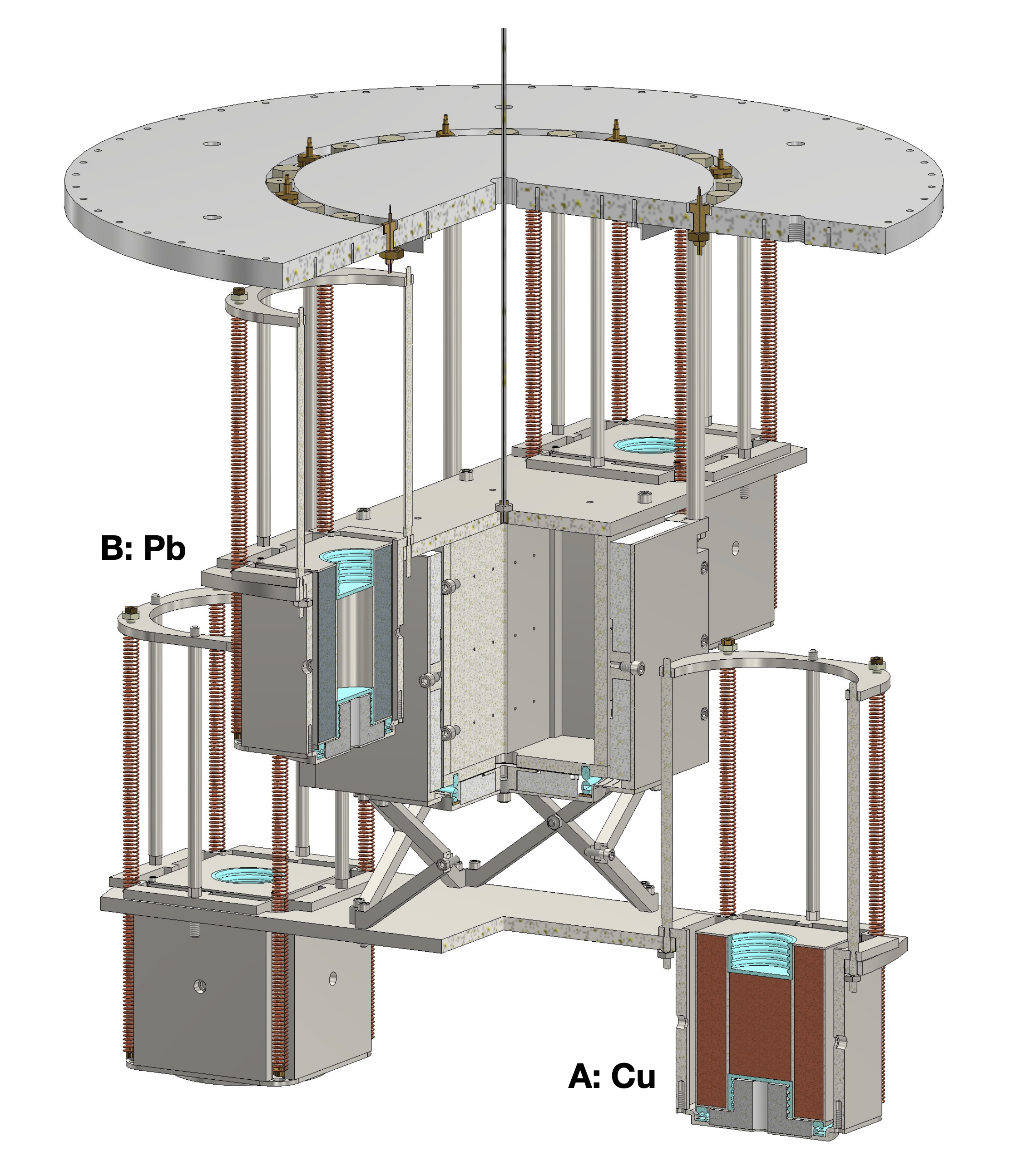}
\caption{\label{fig:composition}The $2+2$ assignment shown on the mSOGRO mechanical design. Each of the four suspended TMs is one of two materials, one lower mass and one upper mass of each kind, labeled A (Cu) and B (Pb); the half-sectioned views show the Cu and Pb inserts caged in identical Ti shells, with the superconducting negative-spring structures of Appendix~\ref{app:tmdesign}. This is the arrangement for which Eq.~(\ref{eq:qC}) puts $\qv$ along $\hat{\bm x}$. Grouping the two lower masses together would put $\qv$ along $\hat{\bm z}$ instead, of the same magnitude, since $\Mmat$ is isotropic. Nothing else in the mechanical design changes: the pattern is imposed by the choice of material for each TM, not by any modification of the suspension, the readout or the geometry.}
\end{figure}

We take one of them as the representative and use it throughout. We assign material $A$ to one upper mass and one lower mass and material $B$ to the remaining two TMs (Fig.~\ref{fig:composition}),
\begin{equation}
\abaralpha_1=\abaralpha_3=\abaralpha_A,
\qquad
\abaralpha_2=\abaralpha_4=\abaralpha_B .
\label{eq:assign}
\end{equation}
The vertex sums that enter Eq.~(\ref{eq:qdef}) are
$\rvec_1+\rvec_3=\tfrac{L}{\sqrt2}\hat{\bm x}$ and
$\rvec_2+\rvec_4=-(\rvec_1+\rvec_3)$, so
\begin{equation}
\begin{aligned}
\qv&=\Mmat^{-1}\left[\abaralpha_A(\rvec_1+\rvec_3)+\abaralpha_B(\rvec_2+\rvec_4)\right]\\
&=\frac{\Delta\abaralpha_{AB}}{L_{\rm eff}}\,\hat{\bm x},
\end{aligned}
\label{eq:qC}
\end{equation}
with $\Delta\abaralpha_{AB}=\abaralpha_A-\abaralpha_B$. The composition vector lies along
$\hat{\bm x}$, the axis separating the two masses of one material from the two
of the other, and its magnitude is the charge contrast divided by an effective
baseline shorter than the arm length by $\sqrt2$. The trace channel then
carries
\begin{equation}
\operatorname{Tr}\K_{\ULDM}=\frac{\Delta\abaralpha_{AB}}{L_{\rm eff}}\,
\gv_\phi(t)\cdot\hat{\bm x},
\label{eq:traceC}
\end{equation}
which is maximal when the field gradient lies along the composition axis and
vanishes when it is perpendicular to it.

\subsection{The $3+1$ assignment}

The remaining way of assigning two materials to four TMs is to give one mass
the second material and the other three the first, which is worth setting against
the $2+2$ choice. Writing
\begin{equation}
\abaralpha_1=\abaralpha_B,
\qquad
\abaralpha_2=\abaralpha_3=\abaralpha_4=\abaralpha_A,
\label{eq:threeone}
\end{equation}
Eq.~(\ref{eq:qdef}) gives
\begin{equation}
\begin{aligned}
\qv&=\Mmat^{-1}\Delta\abaralpha_{BA}\,\rvec_1\\ &=\frac{\Delta\abaralpha_{BA}}{\sqrt2\,L}\left(1,1,1\right)^{\rm T},
\end{aligned}
\label{eq:qthreeone}
\end{equation}
whose magnitude is $\tfrac{\sqrt3}{2}\lvert\Delta\abaralpha_{BA}\rvert/L_{\rm eff}$, smaller than the $2+2$ value by that factor in amplitude and by $\tfrac34$ in power, and whose direction is a cube diagonal rather than a cube axis. The $2+2$ assignment is therefore preferred. It is larger in magnitude, and it uses two masses of each material.

\subsection{The measured tensor}

With the $2+2$ assignment adopted and $\qv$ along $\hat{\bm x}$, the surviving components are
\begin{equation}
\begin{aligned}
&K_{xx}=g_xq_x, && K_{xy}=0,\\
&K_{yx}=g_yq_x, && K_{yy}=0,\\
&K_{zx}=g_zq_x, && K_{zy}=0 ,
\end{aligned}
\label{eq:survivors}
\end{equation}
so the signal occupies the first column alone. Collecting all the surviving
terms of Sec.~\ref{sec:reconstruction}, with the centrifugal term negligible
in the band, the measured tensor is
\begin{equation}
\K_{\rm meas}=
\begin{pmatrix}
\Gamma_{xx}\!+\!g_xq_x & \Gamma_{xy}\!+\!\dot\Omega_z & \Gamma_{xz}\!-\!\dot\Omega_y\\
\Gamma_{xy}\!-\!\dot\Omega_z\!+\!g_yq_x & \Gamma_{yy} & \Gamma_{yz}\!+\!\dot\Omega_x\\
\Gamma_{xz}\!+\!\dot\Omega_y\!+\!g_zq_x & \Gamma_{yz}\!-\!\dot\Omega_x & \Gamma_{zz}
\end{pmatrix}+\mathbf n .
\label{eq:measblock}
\end{equation}
The second and third columns carry no dark matter, since $q_y=q_z=0$. The trace is the one channel gravity leaves empty,
\begin{equation}
\operatorname{Tr}\K_{\rm meas}=g_xq_x+\operatorname{Tr}\mathbf n ,
\label{eq:tracesignal}
\end{equation}
because a gravitational wave is traceless, the Euler term has no trace, and
a gravity gradient in vacuum has $\operatorname{Tr}\bm\Gamma=-4\pi G\rho_{\rm local}=0$. Two antisymmetric combinations are free of gravity as well, since
$\bm\Gamma$ is symmetric and cancels in
\begin{equation}
K_{zx}-K_{xz}=2\dot\Omega_y+g_zq_x,\qquad
K_{zy}-K_{yz}=-2\dot\Omega_x.
\label{eq:cleanantisym}
\end{equation}
The second gives the angular acceleration. The first mixes $\dot\Omega_y$
with the vertical gradient $g_zq_x$ that the trace misses, and the material
exchange $\qv\to-\qv$ separates them, since it reverses $g_zq_x$ and leaves
$\dot\Omega_y$ alone.

Apart from the trace, every entry of Eq.~(\ref{eq:measblock}) that carries the dark-matter signal includes a gravity gradient or an angular acceleration as well. They behave differently in time, however, and that difference separates them. The dark-matter term is a narrow line, of fractional width $Q_{\rm coh}^{-1}\sim10^{-6}$ at a fixed $f_\phi$. The gravitational terms are a static offset, tidal harmonics at frequencies known in advance from the ephemeris, and a broadband continuum. A spectrum therefore separates most of them, the static offset and the centrifugal term lying at zero frequency and outside the band. Whatever remains at the candidate frequency faces two further tests. The three measured rows must be proportional to the same $\qv$, and the line must modulate over a sidereal day, whereas the tidal lines follow the solar and lunar periods. What competes in the remaining bins is the broadband noise floor, against which Fig.~\ref{fig:sens} is computed.

\begin{figure}[t]
\centering
\includegraphics[width=\columnwidth]{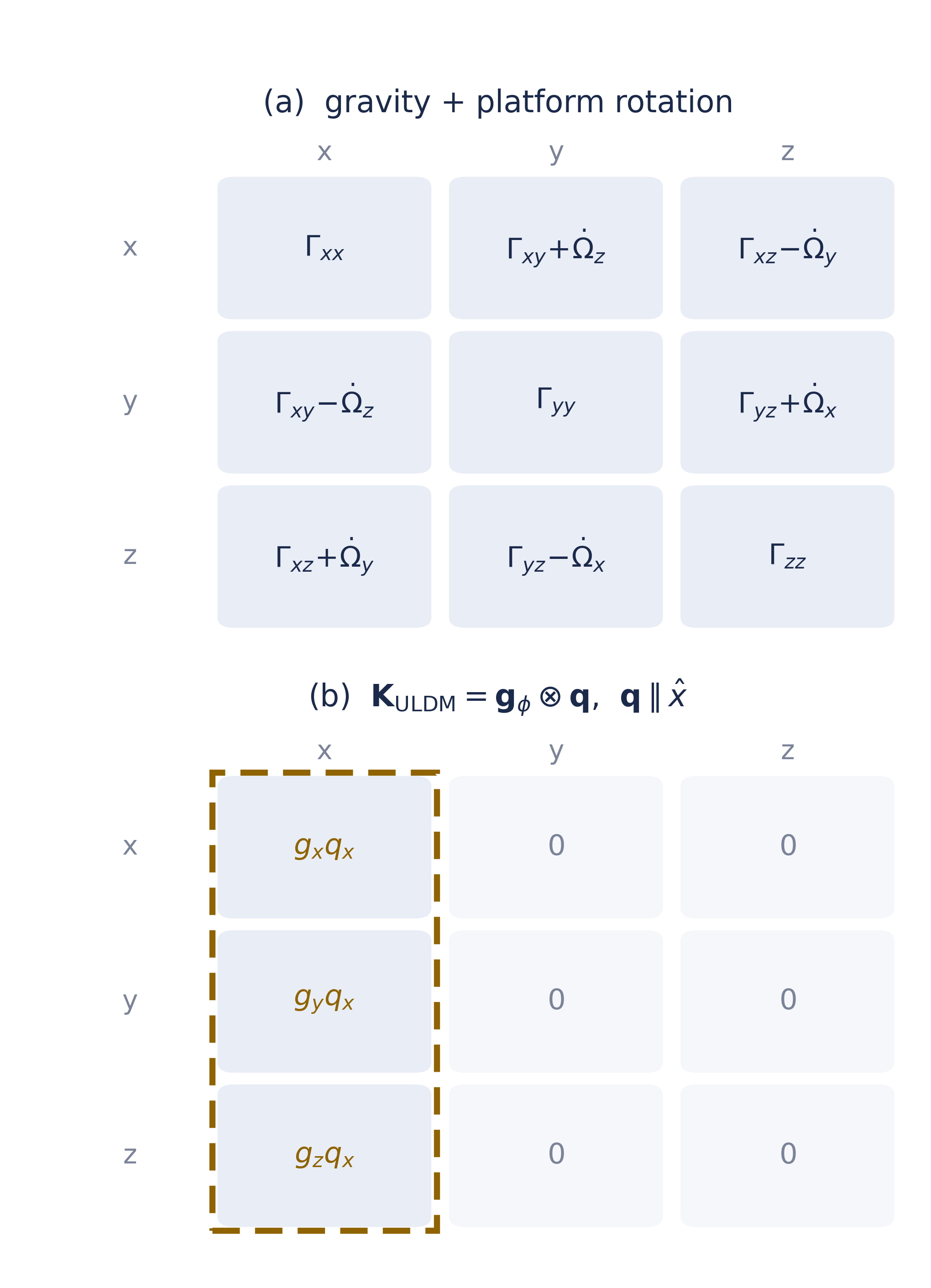}
\caption{\label{fig:channels}How the nine components of the reconstructed tensor are occupied. (a) Gravity contributes the symmetric part and platform rotation the antisymmetric part. (b) The dark-matter response is the outer product $\gv_\phi\otimes\qv$. For the $2+2$ assignment of Sec.~\ref{sec:assignment}, which puts $\qv$ along $\hat{\bm x}$, only the first column survives. The other two splits put $\qv$ along $\hat{\bm y}$ and $\hat{\bm z}$ and fill the second and third columns instead.}
\end{figure}

Keeping only the dark-matter term in Eq.~(\ref{eq:measblock}) leaves
\begin{equation}
\K_{\ULDM}=q_x\begin{pmatrix}g_x&0&0\\ g_y&0&0\\ g_z&0&0\end{pmatrix},
\label{eq:signalpattern}
\end{equation}
so nine channels are fixed by three numbers (Fig.~\ref{fig:channels}). Six of them must vanish, and
the remaining three must be one vector scaled by the same $q_x$. That is the
six conditions of Eq.~(\ref{eq:null}), all of them testable. A candidate
signal is therefore tested by consistency across the channels rather than
by an excess in a single one. Confirmation requires the controls of Sec.~\ref{sec:discussion} and the material exchange.

\section{Sensitivity}\label{sec:sensitivity}

\begin{figure*}[t]
\centering
\includegraphics[width=\textwidth]{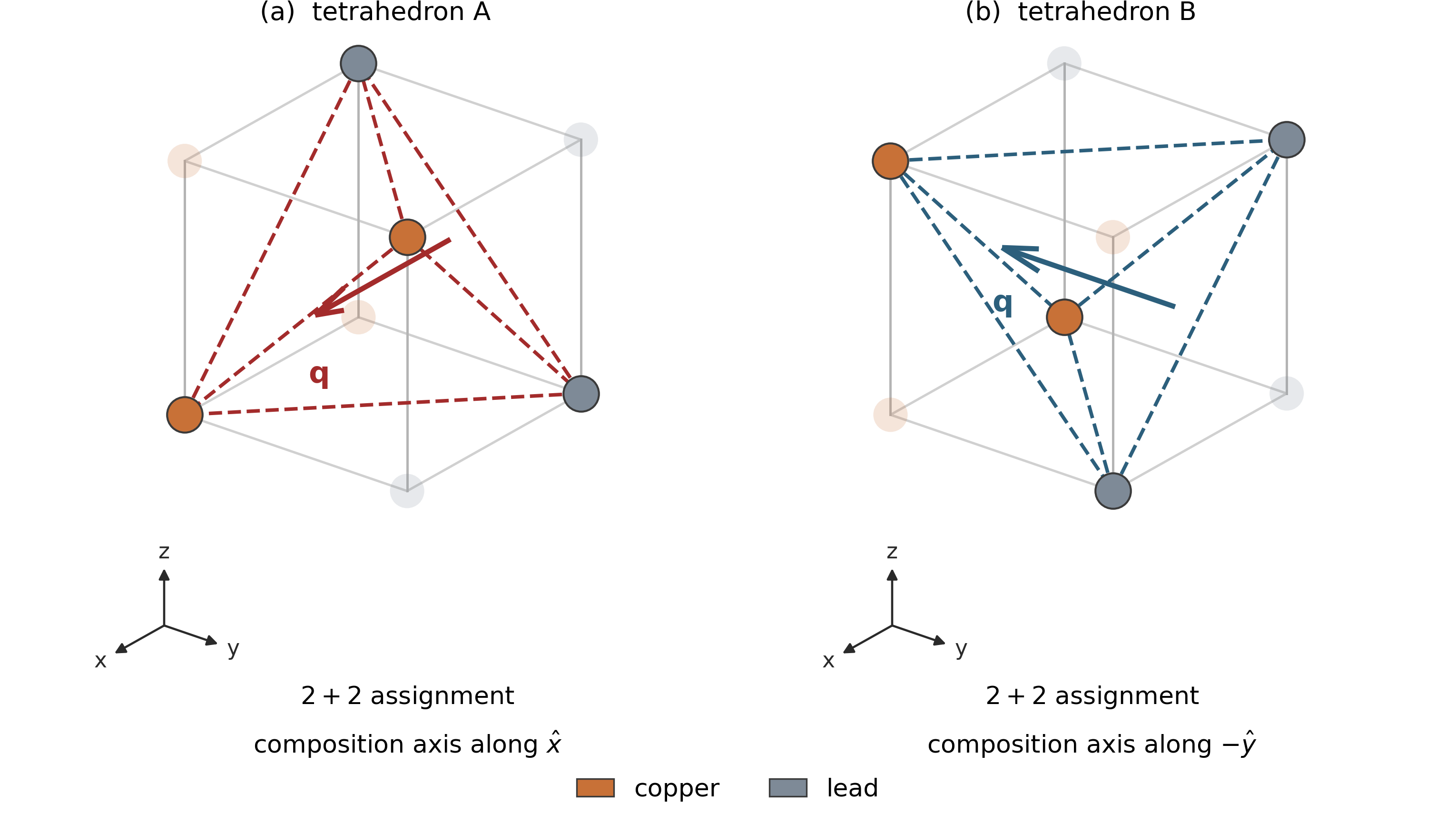}
\caption{\label{fig:twotet}Two tetrahedra interleaved on the vertices of a common rectangular frame; the masses of the other array are shown faintly in each panel. Both use the same copper--lead pair in the $2+2$ assignment of Sec.~\ref{sec:assignment}. The composition axes are chosen orthogonal, along $\hat{\qv}_1=\hat{\bm x}$ in (a) and $\hat{\qv}_2=-\hat{\bm y}$ in (b), so that their trace responses vanish together only for an instantaneously vertical gradient. The ratio of the two amplitudes is set by the unknown gradient direction. Each tetrahedron measures all three rows; the second adds an independent composition axis rather than tensor rank.}
\end{figure*}

\subsection{Gain from a second tetrahedron}

The mSOGRO structure admits two tetrahedra interleaved on the vertices of a common rectangular frame (Fig.~\ref{fig:twotet}). Both arrays use the same material pair in the $2+2$ assignment of Sec.~\ref{sec:assignment}. We write $\qv_1$ and $\qv_2$ for the two composition vectors, the first being the $\qv$ of Eq.~(\ref{eq:qC}).

The two composition axes may be chosen orthogonal without a third material, because $\qv$ reflects the spatial arrangement of the charges through $\sum_i\abaralpha_i\rvec_i$ and not the material pair alone. The second tetrahedron occupies the four remaining corners of the cube, at $\bm s_{1,4}=\tfrac{L}{2\sqrt2}(\pm1,\mp1,+1)$ and $\bm s_{2,3}=\tfrac{L}{2\sqrt2}(\pm1,\pm1,-1)$, and is regular in the same way. Assigning the materials by the same rule, material $A$ on $\{\bm s_1,\bm s_3\}$, gives group sums $\bm s_1+\bm s_3=-\tfrac{L}{\sqrt2}\hat{\bm y}$ and $\bm s_2+\bm s_4=-(\bm s_1+\bm s_3)$, each pair again being a pair of opposite vertices, so that by the same steps
\begin{equation}
\qv_2=-\frac{\Delta\abaralpha_{AB}}{L_{\rm eff}}\,\hat{\bm y},
\label{eq:qprime}
\end{equation}
of the same magnitude as Eq.~(\ref{eq:qC}) and exactly orthogonal to it, since $\qv_1$ lies along $\hat{\bm x}$ and $\qv_2$ along $\hat{\bm y}$. The remaining two splits reverse the signs, so one material pair realizes $\pm\qv_1$ and $\pm\qv_2$ along two orthogonal azimuths, the sign choice made by material exchange. The two arrays therefore project the same gradient onto two orthogonal axes, and their trace responses cannot vanish together unless the gradient is instantaneously vertical.

The gain of adding the second array depends on which noise sources the two arrays share. Because the sensitive axes of the two arrays are orthogonal, correlating their outputs offers a handle on the random seismic and Newtonian noise. The thermal and amplifier noise of the accelerometers is independent and gives $\sqrt2$ in signal-to-noise and $1/\sqrt2$ in the coupling sensitivity. That factor belongs to the optimal joint analysis of the two arrays. The difference statistic of Eq.~(\ref{eq:refmode}) with orthogonal axes has $\lvert\qv_1-\qv_2\rvert=\sqrt2\,\lvert\qv_1\rvert$ and twice the noise power, so it retains the single-array sensitivity while cancelling the common backgrounds. The gain and the cancellation are properties of two different statistics. The same factor follows from doubling the integration time while the field remains coherent, or from doubling the TM.

The pair also admits difference combinations formed in analysis. The gravitational, gravitational-wave and rotational backgrounds are common to the pair and cancel in the difference of the two reconstructed tensors,
\begin{equation}
\K_{(1)}-\K_{(2)}=\gv_\phi\otimes(\qv_1-\qv_2)+(\mathbf n_1-\mathbf n_2)+\epsilon\,\K_{\rm env}+\delta\K_{\rm spat},
\label{eq:refmode}
\end{equation}
with $\epsilon$ the residual transfer-function mismatch and $\delta\K_{\rm spat}$ the spatial residual of the finite baselines. The cancellation of the common background holds in the local linear-gradient approximation; the curvature of the field is sampled at different points by the two parity classes and survives the difference. 

\subsection{Signal-to-noise and integration time}

The signal is an acceleration, $a_{\rm sig}=A_0\Delta\abaralpha$. Its dependence on $\Delta\abaralpha$ is the violation of the equivalence principle: two materials fall differently. What the instrument must do is resolve that acceleration against its noise, so
\begin{equation}
\mathrm{SNR}\sim\frac{A_0\,\Delta\abaralpha}{S_a^{1/2}}\sqrt{\tau_{\rm eff}}
\propto A_0\,\Delta\abaralpha\sqrt{\frac{mQ}{8T\omega_0}}\sqrt{\tau_{\rm eff}},
\label{eq:snr}
\end{equation}
where the second form uses the thermal acceleration noise of the differential mode of a mass pair, $S_a=8k_BT\omega_0/mQ$, twice the single-mass value because the reduced mass is $m/2$~\cite{Paik1996}, and $\tau_{\rm eff}$ is the effective integration time. While the observation is shorter than $\tau_{\rm coh}$, the field keeps a definite phase throughout and $\tau_{\rm eff}=\tau_{\rm int}$, the observation time. For longer observations the phase resets every $\tau_{\rm coh}$, so the record breaks into $N=\tau_{\rm int}/\tau_{\rm coh}$ segments whose signals add incoherently. Averaging them reduces the fluctuation of the noise power by $\sqrt N$, and since the detection statistic is a power, the amplitude sensitivity improves as $N^{1/4}$, leaving $\tau_{\rm eff}=\sqrt{\tau_{\rm coh}\tau_{\rm int}}$~\cite{Manley2021}. The boundary between the two lies at $f_\phi=Q_{\rm coh}/\tau_{\rm int}\simeq30\,$mHz for a year of data, above which the coupling sensitivity improves only as $\tau_{\rm int}^{1/4}$ rather than $\tau_{\rm int}^{1/2}$.

Setting $\mathrm{SNR}=1$, the convention used in comparable accelerometer proposals~\cite{Manley2021}, and writing $\Delta\abaralpha=\sum_c d_c\Delta Q'_c$, with the core fraction of the Maryland-designed TMs folded in explicitly, $\Delta Q'_{\hat m,\rm eff}=0.903\,\Delta Q'_{\hat m}=0.903\times1.910\times10^{-3}=1.72\times10^{-3}$ (Appendix~\ref{app:tmdesign}) since the shell and the niobium hardware are common to the two TMs and cancel in the charge difference gives the sensitivity in the couplings, combined over all three axes,
\begin{equation}
d_c^{\rm max}\propto\frac{1}{\Delta Q'_c}
\sqrt{\frac{8T\omega_0}{3mQ}}\;\frac{1}{\sqrt{\tau_{\rm eff}}} ,
\label{eq:sens}
\end{equation}
from which the upgrade path can be read directly (Table~\ref{tab:scaling}).

The signal is an acceleration and does not change with the TM, while the thermal noise falls as $1/\sqrt{m}$, so a heavier mass raises the signal-to-noise ratio. The array size does not appear at all. The composition vector scales as $\qabs{\qv}\propto1/L$, so $\K_{\ULDM}\propto1/L$ and the differential acceleration across the array, of order $\K_{\ULDM}L$, is independent of $L$. The instrument noise falls as $1/L$ as well, since $S_{\Gamma,\Sigma}\propto1/(mL^2)$ gives $S_{\Gamma,\Sigma}^{1/2}\propto1/(L\sqrt m)$. Equation~(\ref{eq:sensclosed}) carries $L_{\rm eff}\propto L$ in the numerator against that, so the product $L_{\rm eff}S_{\Gamma,\Sigma}^{1/2}$ is independent of $L$ and only the $m^{-1/2}$ survives. In contrast, the tidal tensor of a gravitational wave is set by the source and does not scale with $L$, while the differential acceleration grows with the baseline.

\begin{table*}[t]
\caption{\label{tab:scaling}How the coupling sensitivity of Eq.~(\ref{eq:sens}) responds to the design parameters.}
\begin{ruledtabular}
\begin{tabular}{lll}
Parameter & Dependence of the sensitivity & Remarks\\
\colrule
charge contrast $\Delta Q'_c$ & $(\Delta Q'_c)^{-1}$ & large and practically attractive\\
test mass $m$ & $m^{-1/2}$ & $4\times$ heavier gains $2$\\
quality factor $Q$ & $Q^{-1/2}$ & $10^8\to10^9$ gains $3.2$\\
temperature $T$ & $T^{1/2}$ & $4.2\,$K$\,\to0.1\,$K gains $6.5$\\
resonance $\omega_0$ & $\omega_0^{1/2}$ & $0.03\to0.01\,$Hz gains $1.7$\\
integration time & $\tau_{\rm eff}^{-1/2}$ & $1\to4\,$yr gains $2$ while coherent, $\sqrt2$ beyond\\
array size $L$ & none & signal and noise scale alike\\
\end{tabular}
\end{ruledtabular}
\end{table*}

\begin{figure}[t]
\centering
\includegraphics[width=\columnwidth]{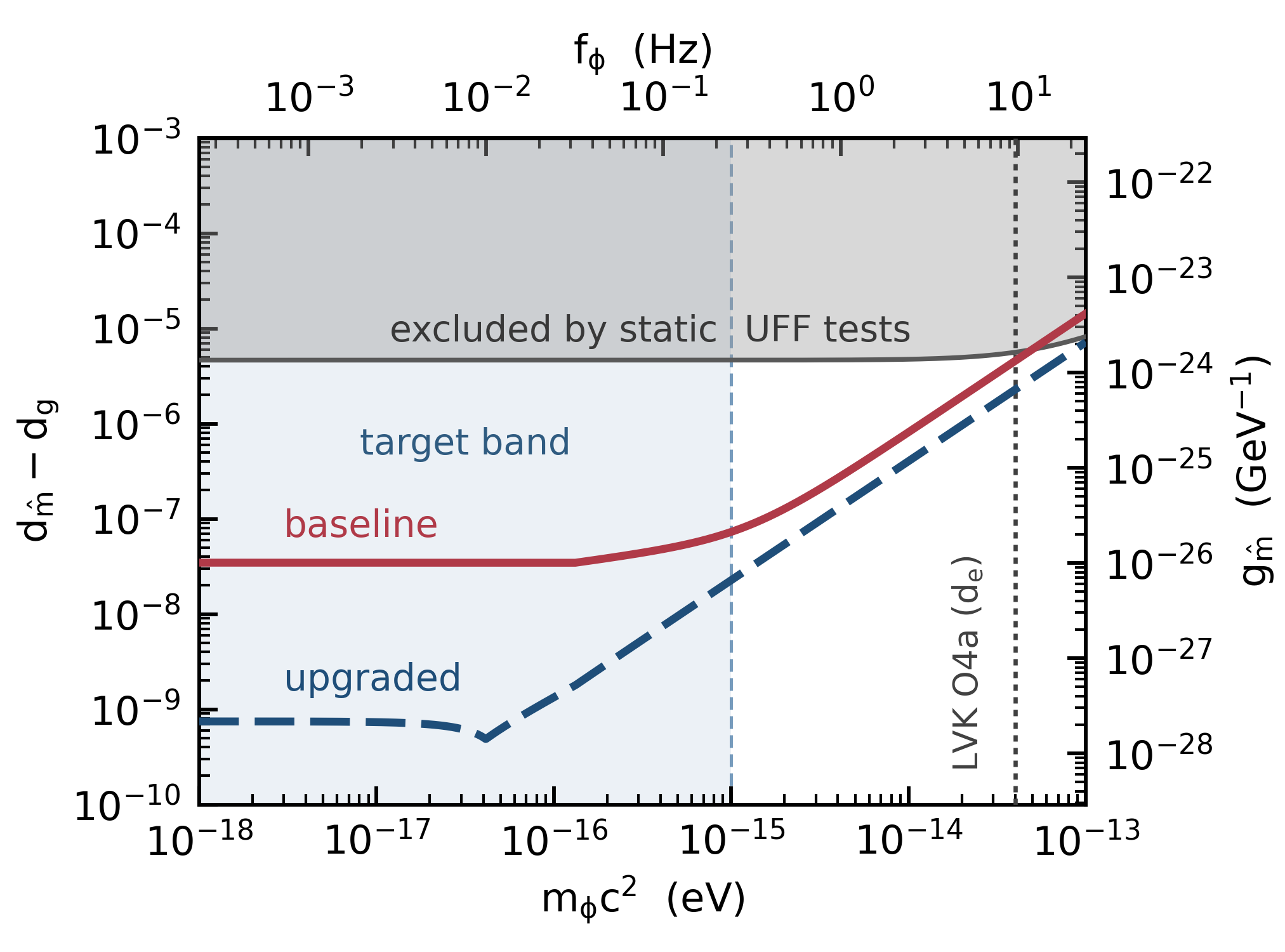}
\caption{\label{fig:sens}Instrument-noise-limited characteristic sensitivity in the dilatonic coupling $d_{\hat m}-d_g$ against the ULDM mass, from Eq.~(\ref{eq:sensclosed}) with the SOGRO instrument noise model~\cite{Paik2016,Paik2020}, for the copper--lead pair in the $2+2$ assignment. Because $A_0$ is independent of $m_\phi$, each curve has the shape of the instrument noise itself, flat where the thermal term dominates and rising where the amplifier term takes over. The right-hand axis restates the sensitivity as $g_{\hat m}=(d_{\hat m}-d_g)\sqrt{4\pi G}$, with $(4\pi G)^{-1/2}=3.44\times10^{18}\,$GeV. The shaded region is excluded by the static tests of Appendix~\ref{app:charges}, the tinted band is the mass window of Eq.~(\ref{eq:fphi}), and the dotted line gives the onset of the interferometer dilaton searches.}

\end{figure}

\subsection{Sensitivity in the dilatonic coupling}

Turning Eq.~(\ref{eq:sens}) into a curve requires a noise model. The gradient-noise power spectral density, combined over the six differential accelerometer channels of the SOGRO noise model, is
\begin{equation}
S_{\Gamma,\Sigma}(f)=\frac{8}{9mL^2}\left[\frac{k_BT\,\omega_D}{Q_D}+\left(1+\beta^{-2}\right)^{1/2}n\hbar\,\lvert\omega^2-\omega_D^2\rvert\right],
\label{eq:noisemodel}
\end{equation}
with Nyquist thermal noise in the first term and the SQUID amplifier noise in the second, following the SOGRO noise modeling~\cite{Paik2016,Paik2020}. The transducer factor $(1+\beta^{-2})^{1/2}$ of that model is unity here, since the design has $\beta\gg1$ below $1\,$Hz. The design parameter values adopted here are those of the mSOGRO accelerometer: $m=10\,$kg, $L=0.4\,$m, $T=4.2\,$K, $f_D=0.03\,$Hz, $Q_D=10^8$ and $n=280$ for the baseline, and $m=40\,$kg, $T=0.1\,$K, $Q_D=10^9$ and $f_D=10\,$mHz for the upgrade. For the baseline design, the thermal term alone gives a gradient noise of $2.46\times10^{-16}\,\mathrm{s^{-2}\,Hz^{-1/2}}$.

The coupling sensitivity follows from setting the signal gradient equal to the noise. By Eq.~(\ref{eq:qC}) the composition response puts a gradient of magnitude $A_0\Delta\abaralpha/L_{\rm eff}$ into the tensor channel. Equating it to $\sqrt{2S_{\Gamma,\Sigma}/\tau_{\rm eff}}$ fills in the proportionality of Eq.~(\ref{eq:sens}) and gives the curve of Fig.~\ref{fig:sens},
\begin{equation}
d^{\max}_{\hat m}(f)=\frac{L_{\rm eff}}{A_0\,\lvert\Delta Q'_{\hat m,\rm eff}\rvert}\sqrt{\frac{2\,S_{\Gamma,\Sigma}(f)}{\tau_{\rm eff}(f)}},
\label{eq:sensclosed}
\end{equation}
where $\tau_{\rm coh}=Q_{\rm coh}/f$. The phase of the line is not known in advance, so the detection statistic is the power in both quadratures, which is the factor of two inside the square root.

In a year the baseline curve is lowest near $30\,$mHz, at $d_{\hat m}-d_g\simeq3.4\times10^{-8}$. Upgrading the accelerometers lowers the curve and moves its minimum to $10\,$mHz, following $f_D$, where it reaches $4.9\times10^{-10}$. The two minima differ by a factor of $70$, which comes entirely from parameters shared with the gravitational-wave channel.

The static tests of Appendix~\ref{app:charges} are the bound to compare with. The interferometer dilaton searches~\cite{LVKO4a2025} constrain photon and electron couplings rather than the combination plotted here, and the clock comparisons of Ref.~\cite{Hees2016} constrain a combination dominated by $d_e$ with a percent-level admixture of $d_{\hat m}-d_g$, up to masses of about $2\times10^{-18}\,$eV. They overlap the low-mass end of the window of Eq.~(\ref{eq:fphi}) but do not reach the band of the sensitivity minima. Figure~\ref{fig:sens} gives the coupling at which the response would reach unit signal-to-noise in the instrument noise, whereas the static bounds are couplings already excluded by measurement.

\section{Summary and discussion}\label{sec:discussion}

The composition-dependent oscillating response of a tetrahedral SGG is derived here, together with the null condition that separates the signal from gravity.

\begin{table}[t]
\caption{\label{tab:steps}From the coherent local gradient to the channel content of the response.}
\begin{ruledtabular}
\begin{tabular}{ll}
Step & Result\\
\colrule
coherent local gradient & $\bm a_i^{\ULDM}(t)=\abaralpha_i\,\gv_\phi(t)$\\
charge assignment & $\abaralpha_i=\abaralpha_0+\qv\cdot\rvec_i$\\
reconstruction, Eq.~(\ref{eq:Kreco}) & $\K_{\ULDM}(t)=\gv_\phi(t)\otimes\qv$\\
null condition & $\K_{\ULDM}(\mathbf I-\Pq)=0$\\
decomposition & $1_{\rm trace}\oplus5_{\STF}\oplus3_{\rm rot}$\\
measured channels & all nine, by Eq.~(\ref{eq:tiltsolved})\\
\end{tabular}
\end{ruledtabular}
\end{table}
The mean composition charge enters the translational common mode and is rejected together with platform acceleration. Only the charge contrast $\Delta\abaralpha_{AB}$ survives, entering the readout as the rank-one tensor $\K_{\ULDM}(t)=\gv_\phi(t)\otimes\qv$, with the axis of $\qv$ fixed in advance by the material assignment. Table~\ref{tab:steps} sets out the six steps that lead there, from the acceleration of a single TM to the channels the response occupies.

In the idealized 3D readout the trace of that tensor is free of the gravitational-wave background and of a gravity gradient in vacuum, since both are traceless. The trace remains open to density fluctuations inside the sensing volume, to instrumental effects, and to any scalar-polarization component of a metric perturbation. An external Newtonian disturbance stays out of the trace only to leading order in the gradient expansion. Its true gradient obeys Laplace's equation at the TMs and is traceless. The tensor reconstructed from four points also samples the curvature of the field. 
The STF and antisymmetric components carry additional information, but they are degenerate with ordinary gravity-gradient and rotational backgrounds respectively.

The pendulum-suspended designs, mSOGRO and the proposed lunar LANGO, are soft in the two horizontal directions, and the vertical mode is softened by the superconducting negative spring already demonstrated in the Model~3 gradiometer. All three rows of $\K$ are then read, the tilts are measured rather than assumed small, and the trace is available. 

With TMs of a single composition $\qv$ vanishes identically, whatever the coupling strength, and to leading order in $\lvert\bm k_\phi\rvert L$ the response collapses into a common mode that is rejected. With two compositions in the $2+2$ assignment the same instrument acquires a dark-matter channel of strength proportional to $\Delta\abaralpha$. The response is concentrated into the one column that $\qv$ selects. Nothing in the sensors, the readout or the reconstruction changes.

Composition-dependent violation of the equivalence principle already underlies a mature class of dark-matter searches, in space-borne differential accelerometers, laboratory torsion balances and dual-species atom interferometers~\cite{Touboul2022,Berge2018,Graham2016}. Those instruments compare two materials along essentially a single axis and give one number. The tetrahedral array gives a tensor instead. A candidate signal is then described not only by its composition dependence but by the three sector norms of Eq.~(\ref{eq:norms}), whose amplitudes follow the rotation of the laboratory frame relative to the halo over a sidereal day.

The vertical negative spring has to be built and characterized at the scale used here (see Appendix~\ref{app:tmdesign}), and the platform that carries two interleaved tetrahedra has to be designed. The detection pipeline will need a likelihood on the stochastic gradient together with a systematics budget.

\begin{acknowledgments}
We thank J.~W. Lee for helpful discussions on ultralight dark matter. JJO was supported by the Basic Science Research Program
through the National Research Foundation of Korea (NRF) funded by the Korea Ministry of Science and ICT (MSIT) (RS-2025-00555178) and the National Institute for Mathematical Sciences under the Primary Research Program (Project No. B26720000) supported by MSIT. HJP was supported by the National Science Foundation of the USA under grant PHY2207757.

The ideas, derivations, and proofs presented here are the authors' own. Claude Opus~5 and Fable~5 (Anthropic) were used to check selected calculations, to write the code that generates Fig.~\ref{fig:sens} from Eqs.~(\ref{eq:noisemodel}), (\ref{eq:sensclosed}) and (\ref{eq:microline}), and to assist with editing the manuscript, and GPT-5.5 and GPT-6.0 (OpenAI) were used as additional cross-checks. The authors directed all such use and independently verified all substantive outputs against their derivations and the cited literature. The authors take full responsibility for the content.
\end{acknowledgments}

\section*{Data availability}
No experimental data were created or analyzed in this study. The sensitivity curves of Fig.~\ref{fig:sens} follow from Eqs.~(\ref{eq:noisemodel}) and (\ref{eq:sensclosed}), and the static-test boundaries from Eq.~(\ref{eq:microline}), with the parameters given in the text and in Appendix~\ref{app:charges}. The script that produces the figure is archived at Zenodo, \href{https://doi.org/10.5281/zenodo.22747613}{doi:10.5281/zenodo.22747613}~\cite{SensitivityCode2026}.

\appendix

\section{Dilatonic charges and the static-test boundaries}\label{app:charges}

\subsection{Charges of the two materials}

The charges $Q_{c,A}$ of Eq.~(\ref{eq:alphaA}) follow from the semi-empirical expressions of Damour and Donoghue~\cite{DamourDonoghue2010}, in the form tabulated by Hees et al.~\cite{Hees2018}. A prime denotes that form throughout. Writing
$\delta\equiv(A-2Z)/A$ for the neutron excess,
\begin{equation}
\begin{aligned}
Q'_{\hat m}&=-\frac{0.036}{A^{1/3}}-0.02\,\delta^{2}
            -\frac{1.4\times10^{-4}Z(Z-1)}{A^{4/3}},\\
Q'_{\delta m}&=1.7\times10^{-3}\,\delta,\\
Q'_{m_e}&=-2.75\times10^{-4}\,\delta,\\
Q'_{e}&=-4.1\times10^{-4}\,\delta
        +\frac{7.7\times10^{-4}Z(Z-1)}{A^{4/3}} ,
\end{aligned}
\label{eq:charges}
\end{equation}
with $A$ the mass number and $Z$ the atomic number. For a compound or alloy the charge is the mass-weighted mean of Eq.~(\ref{eq:abar}). Evaluating Eq.~(\ref{eq:charges}) for the two materials of Sec.~\ref{sec:assignment} gives the values in Table~\ref{tab:charges}, and the difference between them is
\begin{equation}
\begin{aligned}
\lvert\Delta Q'_{\hat m}\rvert&=1.910\times10^{-3}, &
\lvert\Delta Q'_{e}\rvert&=1.656\times10^{-3},\\
\lvert\Delta Q'_{\delta m}\rvert&=2.06\times10^{-4}, &
\lvert\Delta Q'_{m_e}\rvert&=3.33\times10^{-5},
\end{aligned}
\label{eq:contrast}
\end{equation}
which is the input to Fig.~\ref{fig:sens}. These are values for the pure metals. Several requirements bear on the choice of the pair. The charge contrast should
be large, since the sensitivity of Eq.~(\ref{eq:sens}) goes as $1/\Delta Q'_c$.
The densities should not be too different, so that the two cores can be matched
in mass without an awkward disparity in size. And the magnetic response of each
core should be stable and well characterized under the operating conditions,
since a composition-dependent magnetic force has exactly the affine structure
that Eq.~(\ref{eq:null}) admits.

Lead is favored on the last two counts. Below its transition at $7.2\,$K, it is
a nearly ideal diamagnet within the Meissner regime, so its magnetic response
is a fixed number rather than a quantity that varies with field, temperature
and crystallographic texture. It is ductile, casts easily and is uniform in
density, so the center of mass can be placed at the geometric center and the
residual error bounded. Bismuth has a charge contrast larger by $1.6\,\%$ and a
smaller density mismatch, but it is brittle, strongly anisotropic and the most
diamagnetic of the elements, and its susceptibility is neither field- nor
temperature-independent. Copper and lead give
$\lvert\Delta Q'_{\hat m}\rvert=1.910\times10^{-3}$ at a density ratio of
$0.79$.

The core is enclosed in a shell of Ti--6Al--4V. Its superconducting transition is measured at $5.1\,$K, depends on heat treatment, and in treated samples does not always reach zero resistance~\cite{Ridgeon2017}. A fully superconducting shell would set the magnetic response of both test masses by the common shell rather than by the core. This is a design principle, not a demonstrated exclusion. A residual composition-dependent magnetic force has exactly the affine structure that Eq.~(\ref{eq:null}) admits. Its upper bound must be established in situ, by the shielding, the continuous magnetometer monitoring and the injected gradients listed as controls in the Letter.

\subsection{Construction of the boundaries in Fig.~\ref{fig:sens}}\label{app:static}
The shaded region of Fig.~\ref{fig:sens} is bounded by two static tests of the universality of free fall. Both boundaries are drawn from published E\"otv\"os parameters through the same construction, recorded here for traceability. Equation~(40) of~\cite{Hees2018} gives the E\"otv\"os parameter for a linear coupling as $\eta=\Delta\tilde\alpha\,s_\oplus\,(1+r/\lambda)e^{-r/\lambda}$, with the scalar charge of the source $s_\oplus=\tilde\alpha_\oplus I(R_\oplus/\lambda)$ from its Eq.~(21). Taking one coupling nonzero at a time puts $\Delta\tilde\alpha=\Delta Q'_{\hat m}(d_{\hat m}-d_g)$ on the TMs and $\tilde\alpha_\oplus=Q_{\hat m,\oplus}(d_{\hat m}-d_g)$ on the source, so the coupling enters as a square,
\begin{equation}
\lvert\eta\rvert=Q_{\hat m,\oplus}\,\lvert\Delta Q'_{\hat m}\rvert\,\left(d_{\hat m}-d_g\right)^{2}S(\lambda).
\label{eq:etaform}
\end{equation}
Since Eq.~(\ref{eq:etaform}) is monotonic in the coupling, a measurement bounding $\eta$ bounds the coupling. Writing
\begin{equation}
d^{\rm stat}_{\hat m}\equiv\sqrt{\frac{\lvert\eta\rvert_{95}}{Q_{\hat m,\oplus}\,\lvert\Delta Q'_{\hat m}\rvert\,S(\lambda)}}\,,
\label{eq:microline}
\end{equation}
$\lvert\eta\rvert\le\lvert\eta\rvert_{95}$ gives $\lvert d_{\hat m}-d_g\rvert\le d^{\rm stat}_{\hat m}$, the boundary of the shaded region of Fig.~\ref{fig:sens}, with
\begin{equation}
\begin{aligned}
S(\lambda)&=\Phi(x)\left(1+\frac{r}{\lambda}\right)e^{-r/\lambda},\\
\Phi(x)&=\frac{3(x\cosh x-\sinh x)}{x^{3}},
\end{aligned}
\label{eq:Sfactor}
\end{equation}
where $x=R_\oplus/\lambda$, $R_\oplus$ the Earth's radius, $\lambda=\hbar/(m_\phi c)$ and $r$ the distance from the Earth's center. The factor $\Phi$ is the form factor $I$ of Eq.~(21) of~\cite{Hees2018} and the rest is the radial dependence of its Eq.~(40). Equation~(\ref{eq:microline}) is the inversion of Eq.~(\ref{eq:etaform}) and is assembled here.

For MICROSCOPE, $r=R_\oplus+710\,$km is the orbital radius, $\lvert\eta\rvert_{95}=5.4\times10^{-15}$ derives from~\cite{Touboul2022} with the uncertainties combined in quadrature about zero, a processing choice made here, and $\lvert\Delta Q'_{\hat m}({\rm Ti,Pt})\rvert=2.69\times10^{-3}$ is from the tables of~\cite{Hees2018}. For the E\"ot-Wash rotating balance the source distance is set to the Earth's surface, $\lvert\eta\rvert_{95}=3.5\times10^{-13}$ follows from $\eta({\rm Be,Ti})=(0.3\pm1.8)\times10^{-13}$~\cite{Schlamminger2008,Wagner2012} and $\lvert\Delta Q'_{\hat m}({\rm Be,Ti})\rvert=7.2\times10^{-3}$ from Eq.~(\ref{eq:charges}), giving a flat level of $2.3\times10^{-5}$. Both bounds take one coupling nonzero at a time. The same construction with the $d_e$ charges of the test pairs and the Earth gives the boundary in $d_e$, not drawn in Fig.~\ref{fig:sens}. The two kinds of statement should not be read as one. The static bounds are sourced by the Earth rather than by the halo and hold whatever fraction of the dark matter the scalar makes up, whereas the sensitivity curves of Fig.~\ref{fig:sens} assume that the scalar makes up the whole of the local dark matter, which normalizes $A_0$. Over most of the mass window the curves lie two to four orders of magnitude below the static bound.

\begin{table}[t]
\squeezetable
\caption{\label{tab:charges}Dilatonic charges of the two materials, from Eq.~(\ref{eq:charges}). All entries are in units of $10^{-3}$.}
\begin{ruledtabular}
\begin{tabular}{lcccc}
& $-Q'_{\hat m}$ & $Q'_{e}$ & $Q'_{\delta m}$ & $-Q'_{m_e}$\\
\colrule
copper & $9.622$ & $2.430$ & $0.1485$ & $0.0240$\\
lead & $7.712$ & $4.086$ & $0.3544$ & $0.0573$\\
\end{tabular}
\end{ruledtabular}
\end{table}

\section{Tensor differential accelerometer for ULDM search}\label{app:tmdesign}

\subsection{Test masses with large ULDM charge contrast}

To search for ULDM, we will modify the tetrahedral mSOGRO~\cite{Metzler2022} by constructing two TMs with one material and the other two with another material, as shown in Fig.~\ref{fig:tetrahedron}(b). The TMs of a sensitive ULDM detector must satisfy the following requirements:
\begin{enumerate}
\item To maximize the differential acceleration signal due to ULDM, TM materials with a large ULDM charge contrast must be chosen.
\item To minimize the thermal noise of the TMs for the given baseline ($0.40\,$m), the TMs must be constructed with high-density materials.
\item To be able to couple to the same transducer circuit elements with matched parasitic forces, the outer surfaces of the TMs must be matched.
\item To match the dynamics of the TMs, all the TMs must have equal mass and an equal center-of-mass position with respect to the outer surfaces.
\end{enumerate}
To satisfy requirements (1) and (2), we chose Cu (density $8.94\,\mathrm{g\,cm^{-3}}$, $Q'_{\hat m}=-9.622\times10^{-3}$) and Pb (density $11.34\,\mathrm{g\,cm^{-3}}$, $Q'_{\hat m}=-7.712\times10^{-3}$) as the main components of TM~A and B. To satisfy requirements (3) and (4), we designed the Cu and Pb inserts with equal mass, although of different shapes. They will be caged in a Ti-alloy (Ti--6Al--4V) shell of identical geometry to which niobium (Nb) negative spring structures of identical geometry are attached. This Ti alloy combines high strength, stiffness and low density ($4.43\,\mathrm{g\,cm^{-3}}$), and is easy to machine.

\begin{figure}[!htb]
\centering
\includegraphics[width=\columnwidth]{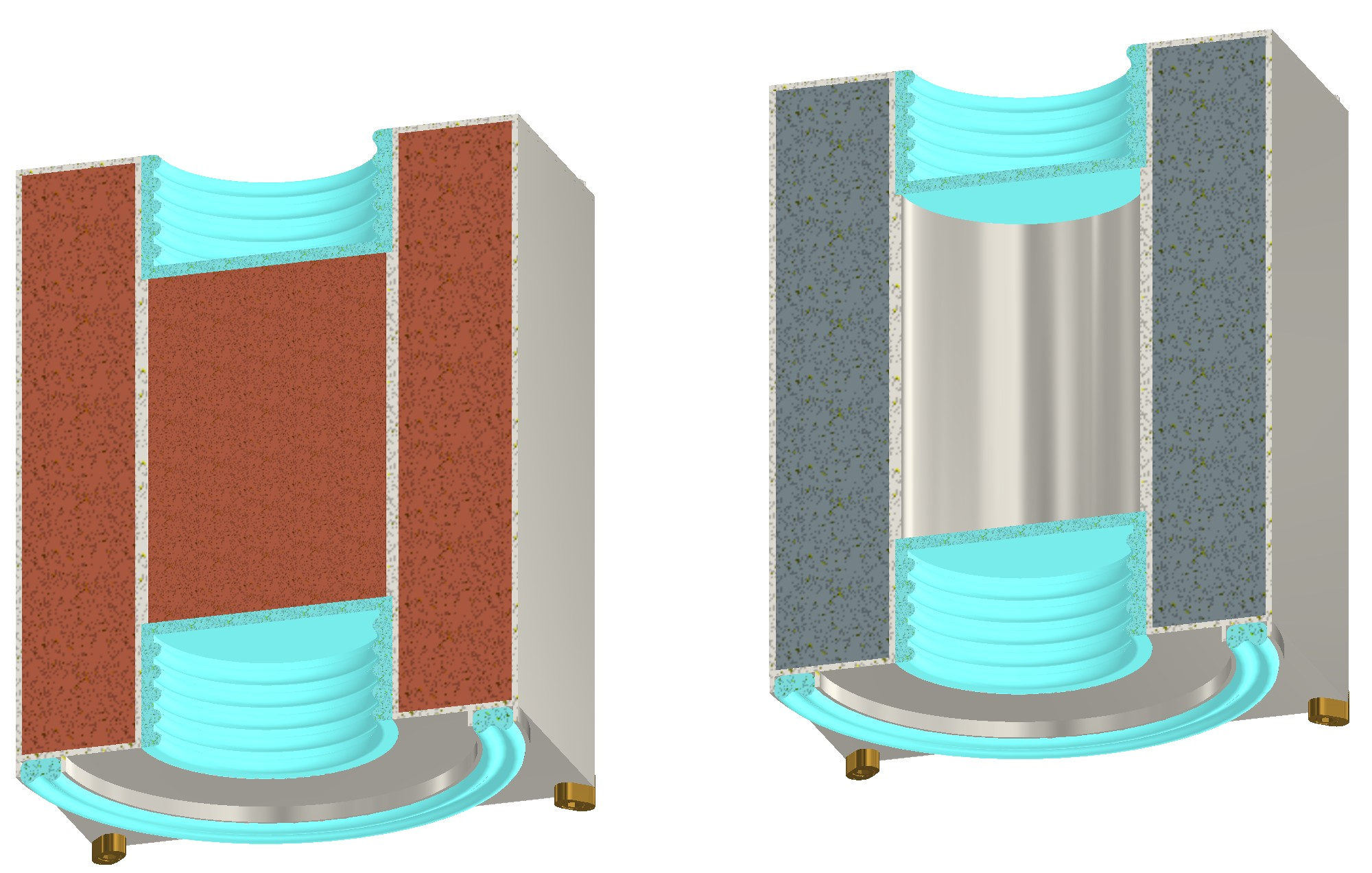}
\caption{\label{fig:tmpair}Half-section views of (left) TM~A with Cu caged in a Ti shell and (right) TM~B with Pb caged in an identical Ti shell. Each TM weighs $10\,$kg, of which $9.0\,$kg is for copper or lead and $1.0\,$kg for the Ti shell loaded with Nb inserts and a Nb ring for superconducting negative springs.}
\end{figure}

Figure~\ref{fig:tmpair} shows half-section views of TM~A and TM~B, from a perspective that shows the bottom surfaces. The outer dimensions of the fully assembled TMs are $10.2\,\mathrm{cm}\times10.2\,\mathrm{cm}\times12.1\,\mathrm{cm}$ (excluding the protrusion at the bottom for a Nb ring for horizontal negative spring) and the thickness of the Ti shell is $1.27\,$mm. The design gives a total mass of $10\,$kg to each TM, with $9.03\,$kg each for the Cu and Pb insert, and $0.97\,$kg for the Ti shell loaded with the Nb and brass components. The Cu and Pb inserts were designed to have a reflection symmetry about the coordinate planes with their origin at the respective geometric centers. Therefore, their centers of mass coincide with their geometric centers. The slightly asymmetrically loaded Ti shell shifts the centers of mass of the total TMs by a small but the same distance.

\subsection{Pendulum suspension and superconducting negative spring}

To obtain high sensitivity for acceleration or gravity sensors, it is essential that the TM suspension frequencies are lowered to below the signal frequency band. To this end, we suspend each TM with four springs as pendula, supplemented with superconducting anti-springs in both horizontal and vertical directions.

\begin{figure}[!htb]
\centering
\includegraphics[width=0.85\columnwidth]{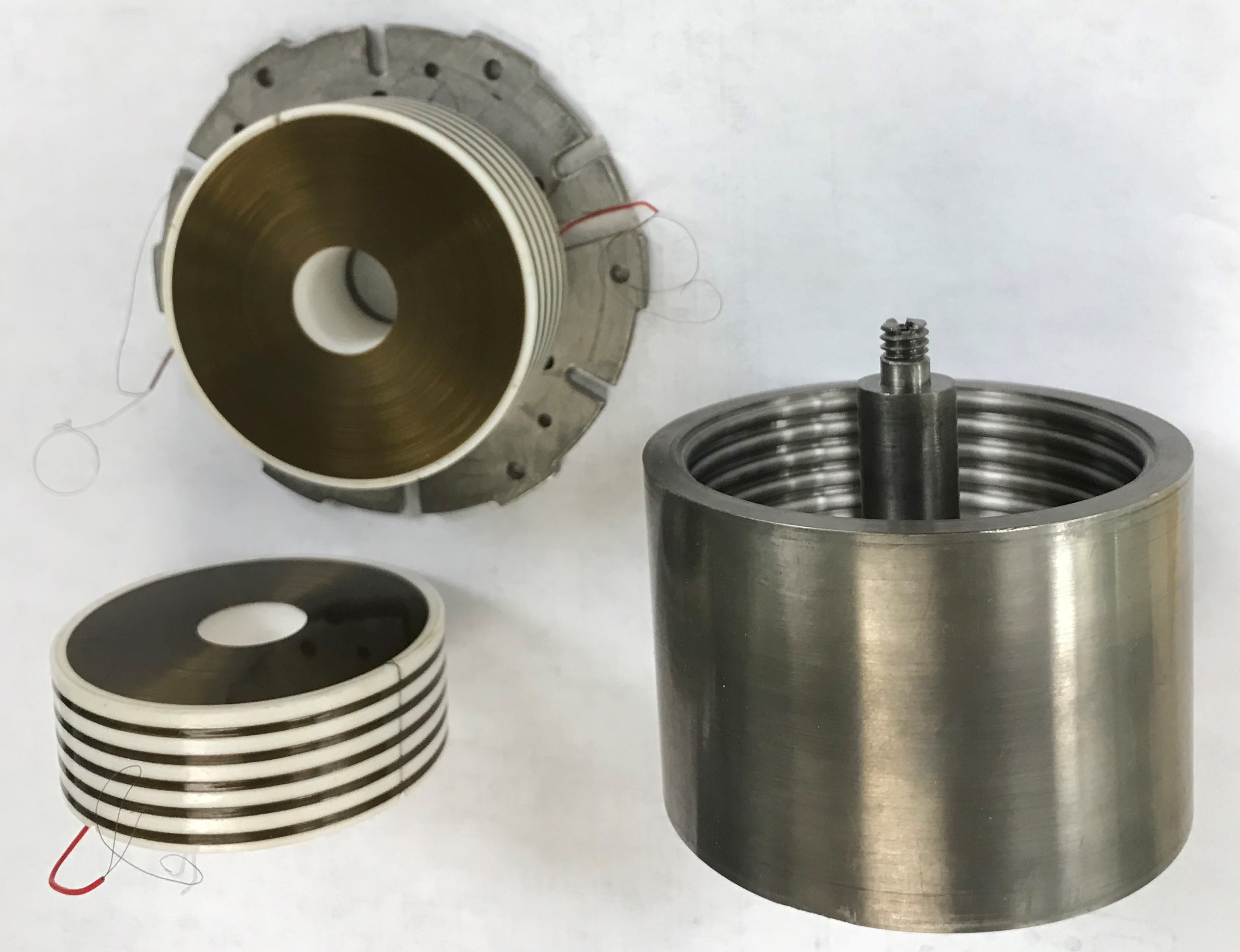}
\caption{\label{fig:nsdemo}Model-3 SGG TM with ridges and matching negative spring coils.}
\end{figure}

By surrounding a superconducting ring with another superconducting ring on its equatorial plane and applying a magnetic field to the surface of the outer ring by storing a persistent current in the inner ring, one can create an unstable equilibrium in the vertical direction, therefore, a \textit{vertical negative spring} for the rings~\cite{MoodyChanPaik1986}. In practice, this can also be achieved with a series of ridges aligned with corresponding coils. Figure~\ref{fig:nsdemo} shows the hardware with which the Maryland group demonstrated the negative spring. A Nb TM had circular ridges, 5 on each side, on the inner surface of its cylindrical shell and matching Nb coils were inserted into the shell. The TM stiffness was reduced from $1440\,\mathrm{N\,m^{-1}}$ to near zero~\cite{Paik1996}. The present experiment requires a vertical negative spring of about $1/3$ of this spring constant.

\begin{figure}[!htb]
\centering
\includegraphics[width=0.6\columnwidth]{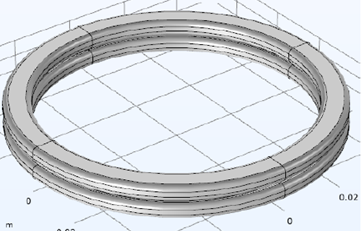}
\caption{\label{fig:hns}Two Nb rings separated vertically to form a horizontal negative spring.}
\end{figure}

To reduce the horizontal resonance frequencies of the TMs from the pendulum frequency $0.94\,$Hz to near zero, we need to design a \textit{horizontal negative spring} with $k_n=-350\,\mathrm{N\,m^{-1}}$. We can achieve a 2D horizontal negative spring by separating the circular rings vertically, as shown in Fig.~\ref{fig:hns}.

\begin{figure}[!htb]
\centering
\includegraphics[width=0.55\columnwidth]{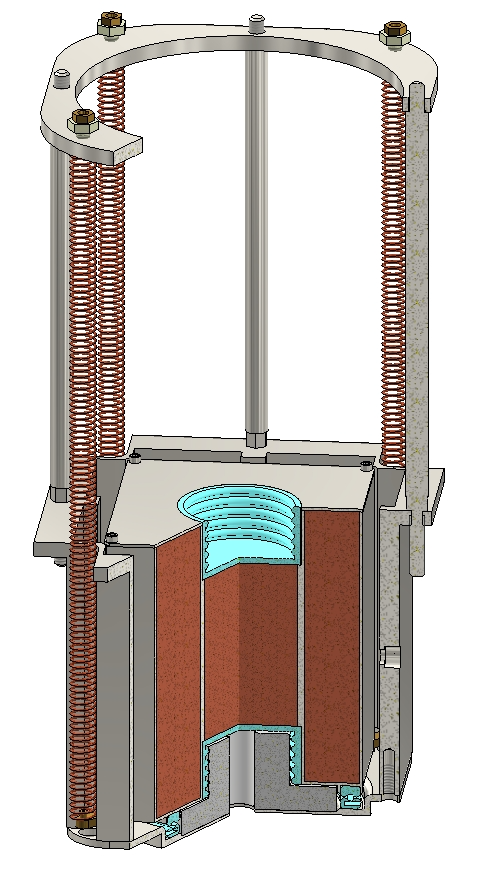}
\caption{\label{fig:suspension}Three-quarter-section view of TM~A (Cu) suspended by four Be-Cu springs. Nb rings (blue) for vertical negative spring are shown at the top, and Nb rings (blue) for horizontal negative spring are shown at the bottom, along with Macor coil form (gray).}
\end{figure}

Figure~\ref{fig:suspension} shows a three-quarter section view of our TM suspended by four beryllium--copper springs. The top TM housing cover is removed to show the Nb vertical negative spring structure (blue). Nb rings (blue) for the horizontal negative spring are shown at the bottom. By suspending the TMs from their bottom surfaces, we obtain a pendulum length of $\ell=28\,$cm, which gives a horizontal resonance frequency of $0.94\,$Hz. Due to the limited space available inside the cryostat, we limit the length of the springs to $12\,$cm and the stretch under the weight of the TM to $14\,$cm. To achieve this, we employ pretensioned beryllium--copper springs. For our design, we obtain a vertical resonance frequency of $1.1\,$Hz. To reduce this resonance frequency to near zero, we need a vertical negative spring with $k_n=-500\,\mathrm{N\,m^{-1}}$.

Each horizontal negative spring ring has two ridges to double the negative stiffness. The bottom ring has a cavity in which a multi-turn coil ($N$ turns with current $i_n$) is wound to generate the field. Our FEM analysis shows that, with total current $Ni_n=580\,$A circulating around the bottom ring, the TM frequency can be nulled without exceeding the critical field of Nb ($h_{c1}=0.125\,$T at $4.2\,$K).

\subsection{LC-bridge sensing circuit}

The tiny mechanical energy that an ULDM flux deposits into the TMs must be converted into an electrical signal and coupled to a nearly quantum-limited SQUID. Since nearly quantum-limited dc SQUIDs have a $1/f$ noise below several kHz, a parametric transducer must be employed to up-convert the signal frequency to above the $1/f$ noise corner frequency. We adopt the mSOGRO approach of employing a capacitor-bridge transducer coupled to two-stage low-noise dc SQUIDs.

\begin{figure}[!htb]
\centering
\includegraphics[width=\columnwidth]{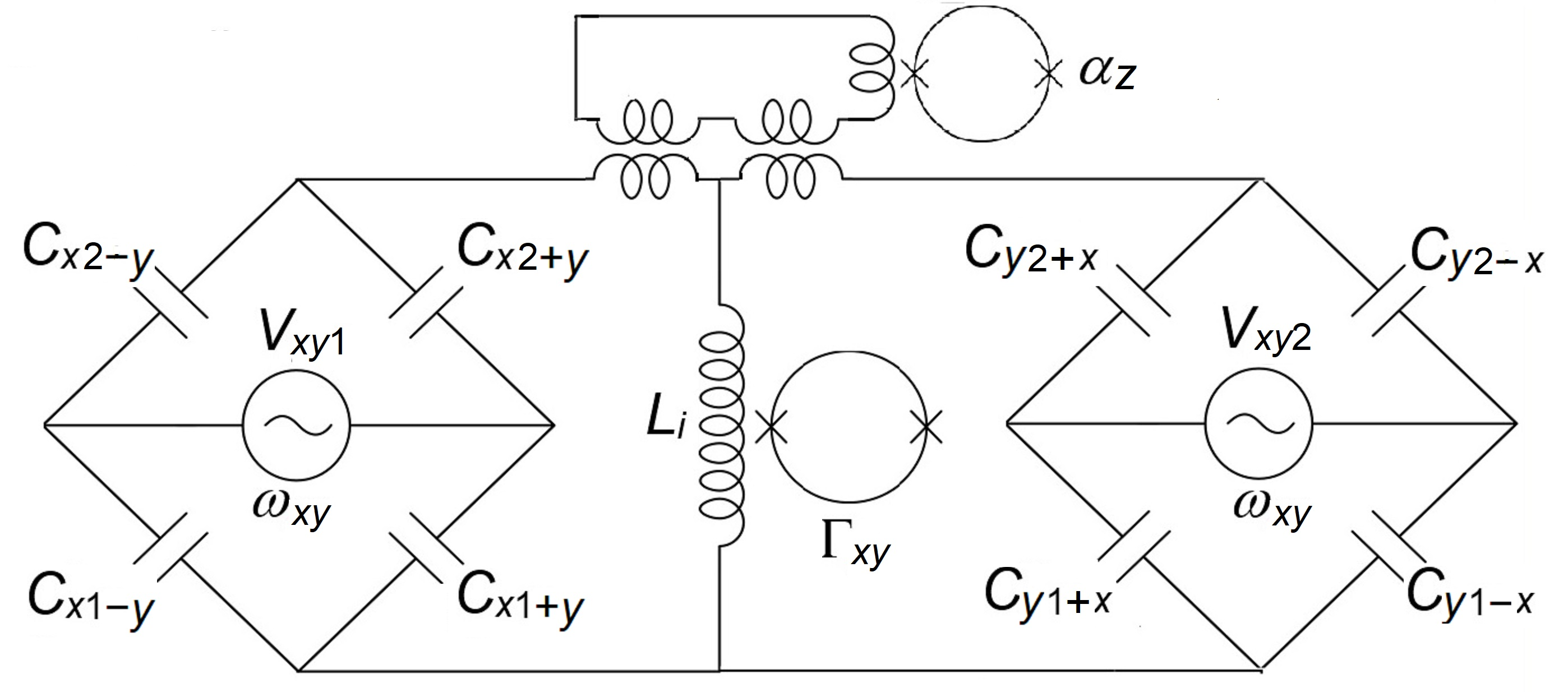}
\caption{\label{fig:lcbridge}Two capacitor bridges are coupled to two low-noise dc SQUIDs, which sum and difference the acceleration signals to detect the CM acceleration and gravity gradient.}
\end{figure}

Figure~\ref{fig:lcbridge} shows a schematic transducer circuit for the $xy$ channel. The sensing capacitors are labeled as follows: $C_{x1+y}$ represents the capacitor for the TM$_1$ on the $x$ axis located on the $+y$ side. Two capacitor bridges are coupled to two low-noise dc SQUIDs, which sum and difference the angular acceleration signals from the two rotating arms to detect CM acceleration $\alpha_z$ and gravity gradient $\Gamma_{xy}$. The sensing capacitances and the inductances at the SQUID input form a superconducting resonant circuit. The capacitor bridge is driven at the LC resonance frequency, $\omega_{xy}=\omega_{LC}$, well above the $1/f$ noise corner frequency of the SQUID. The CM is balanced by adjusting the ratio of the driving voltages of the two capacitor bridges, $V_{xy2}/V_{xy1}$. The sensing circuits for the other tensor components are the same as in Fig.~\ref{fig:lcbridge}, except that the sensing capacitors there are replaced by the capacitors that respond to the particular tensor component.

\par\bigskip
\bibliographystyle{apsrev4-2}
\vspace*{-18pt}
\bibliography{sgg_uldm_theory}

\end{document}